\documentclass[twocolumn,aps,superscriptaddress]{revtex4}
\usepackage{amssymb}
\usepackage{amsmath}
\usepackage{graphicx}
\usepackage[normalem]{ulem}
\usepackage[dvips]{color}
\usepackage{appendix}

\begin{document}

\title{3-dimensional QCD phase diagram with pion condensate in the NJL model}
\author{Lu-Meng Liu}
\affiliation{School of Physical Sciences, University of Chinese Academy of Sciences, Beijing 100049, China}
\author{Jun Xu\footnote{Corresponding author: xujun@zjlab.org.cn}}
\affiliation{Shanghai Advanced Research Institute, Chinese Academy of Sciences, Shanghai 201210, China}
\affiliation{Shanghai Institute of Applied Physics, Chinese Academy of Sciences, Shanghai 201800, China}
\author{Guang-Xiong Peng}
\affiliation{School of Nuclear Science and Technology, University of Chinese Academy of Sciences, Beijing 100049, China}
\affiliation{Theoretical Physics Center for Science Facilities, Institute of High Energy Physics, Beijing 100049, China}
\affiliation{Synergetic Innovation Center for Quantum Effects $\&$ Applications, Hunan Normal University, Changsha 410081, China}
\date{\today}

\begin{abstract}
With the isovector coupling constants adjusted to reproduce the physical pion mass and lattice QCD results in baryon-free quark matter, we have carried out rigourous calculations for the pion condensate in the 3-flavor Nambu-Jona-Lasinio model, and studied the 3-dimensional QCD phase diagram. With the increasing isospin chemical potential $\mu_I$, we have observed two nonzero solutions of the pion condensate at finite baryon chemical potentials $\mu_B$, representing respectively the pion superfluid phase and the Sarma phase, and their appearance and disappearance correspond to a second-order (first-order) phase transition at higher (lower) temperatures $T$ and lower (higher) $\mu_B$. Calculations by assuming equal constituent mass of $u$ and $d$ quarks would lead to large errors of the QCD phase diagram within $\mu_B \in (500, 900)$ MeV, and affect the position of the critical end point.
\end{abstract}

\maketitle

\section{INTRODUCTION}
\label{introduction}

Understanding the structure of the phase diagram for the Quantum Chromodynamics (QCD) is one of the main goals of high-energy nuclear physics. Efforts are mostly devoted to exploring the 2-dimensional QCD phase diagram, i.e., in the plane of the temperature and the baryon chemical potential. Due to the sign problem~\cite{Kar02,Mur03,Ber05,Aok06,Baz12a,Bed18}, lattice QCD (LQCD) calculations are unable to provide solid information in the high baryon chemical potential region, where our knowledge on the QCD phase diagram mostly relies on low-energy relativistic heavy-ion collisions and effective QCD models. Experiments at RHIC-BES, FAIR-CBM, and also at NICA and HIAF, etc., have been or are to be carried out, in order to search for the signal of the QCD critical end point, which was observed in effective QCD models, such as the Nambu-Jona-Lasinio (NJL) model~\cite{Asa89,Fuk08,Car10,Bra13}, the Dyson-Schwinger approach~\cite{Xin14,Fis14}, and the functional renormalization group method~\cite{Fu19,Gao20}.

Our knowledge on the QCD phase diagram can be extended to another degree of freedom, i.e., the isospin~\cite{Son01}. As the isospin chemical potential increases and reaches the mass of a pion, pions can be produced out of the vacuum and a Bose-Einstein condensate is expected to form. The formation of the pion condensate has profound effects on the QCD phase diagram~\cite{Kle03,Bar04,HeLY-ZhuangPF2005,Bar05,Ebe06a,Zha07,Zha07b,Sas10,Fu10,Xia13,Adh18,Bra18a,Khu19,Wu21}. Unlike the case at finite baryon chemical potentials, LQCD calculations do not suffer from the sign problem and can give reliable results at finite isospin chemical potentials. Intuitively, the QCD phase diagram at finite isospin chemical potentials is related to the isovector interaction in quark matter, and the latter has ramifications in both relativistic heavy-ion collisions and nuclear astrophysics. For example, in relativistic heavy-ion collisions induced by neutron-rich nuclei at RHIC-BES, the elliptic flow splitting between $\pi^-$ and $\pi^+$ favors a finite isovector interaction in quark matter~\cite{Liu19}, and the charge susceptibility is largely affected by the isovector interaction~\cite{Liu20a}. The isovector quark interaction may also affect properties of strange quark stars~\cite{Chu15,Liu20}. It is of great interest to reproduce the LQCD results~\cite{Ava19,Lop21} at finite isospin chemical potentials but zero baryon chemical potential by varying the strength of the isovector quark interaction, and then extrapolate the calculations to finite baryon chemical potentials, thus exploring the whole 3-dimensional QCD phase diagram.

In this manuscript, we report such a study based on a 3-flavor NJL model. We reproduce the physical pion mass and LQCD results in baryon-free quark matter by adjusting the coupling constants of the scalar-isovector and vector-isovector interaction. Afterwards, we extrapolate the model to finite baryon chemical potentials, and study on the 3-dimensional QCD phase diagram in the presence of the pion condensate. We note that the constituent masses of $u$ and $d$ quarks are generally set to be equal in previous studies at small or vanishing baryon chemical potentials (see, e.g., Refs.~\cite{Xia13,Lop21,Cao21}). In the present study for exploring the whole 3-dimensional QCD phase diagram, we carry out a rigourous calculation without this drawback. Effects from productions of other mesons, e.g., the kaon condensate, are neglected in the present study. Section~\ref{theory} gives the main formulas for the 3-flavor NJL model, with the derivations of the Lagrangian in the mean-field approximation, the thermodynamic potential, and the quark condensates and densities detailed in the Appendices. Results in baryon-free and baryon-rich quark matter as well as the 3-dimentional QCD phase diagram are given in Sec.~\ref{results}. We conclude and outlook in Sec.~\ref{summary}.

\section{Theoretical framework of 3-flavor NJL model}
\label{theory}

\subsection{The Lagrangian}

We start from the Lagrangian density of a 3-flavor NJL model expressed as~\cite{Liu16}
\begin{equation} \label{NJL Lagrangian}
	\mathcal{L}_\mathrm{NJL}
	=\mathcal{L}_0 + \mathcal{L}_S + \mathcal{L}_V + \mathcal{L}_{\text{KMT}} + \mathcal{L}_{IS} + \mathcal{L}_{IV}, \\
\end{equation}
where
\begin{eqnarray}
	\mathcal{L}_0&=&\bar{\psi}(i\gamma^\mu \partial_\mu - \hat{m})\psi, \\
	\mathcal{L}_S&=& \frac{G_{S}}{2}\sum_{a=0}^8 [(\bar{\psi}\lambda^a\psi)^2 + (\bar{\psi}i \gamma_5 \lambda^a \psi)^2], \\
	\mathcal{L}_V&=& - \frac{G_{V}}{2}\sum_{a=0}^8 [(\bar{\psi}\gamma^{\mu}\lambda^a\psi)^2 + (\bar{\psi} i\gamma_5 \gamma^{\mu} \lambda^a \psi)^2], \\
	\mathcal{L}_{\text{KMT}}&=& -K[\text{det}\bar{\psi}(1+\gamma_5)\psi + \text{det}\bar{\psi}(1-\gamma_5)\psi], \\
	\mathcal{L}_{IS}&=& G_{IS} \sum_{a=1}^3[(\bar{\psi}\lambda^a\psi)^2 + (\bar{\psi}i \gamma_5 \lambda^a \psi)^2], \\
	\mathcal{L}_{IV}&=& - G_{IV} \sum_{a=1}^3[(\bar{\psi}\gamma^{\mu}\lambda^a\psi)^2 + (\bar{\psi} i\gamma_5 \gamma^{\mu} \lambda^a \psi)^2],
\end{eqnarray}
are the kinetic term, the scalar-isoscalar term, the vector-isoscalar term, the Kobayashi-Maskawa-t' Hooft (KMT) term, the scalar-isovector term, and the vector-isovector term, respectively. In the above, $\psi = (u,d,s)^T$ represents the 3-flavor quark fields with each flavor containing quark fields of 3 colors; $\hat{m}=\text{diag}(m_u,m_d,m_s)$ is the current quark mass matrix for $u$, $d$, and $s$ quarks; $\lambda^a $ ($a=1,...,8$) are the Gell-Mann matrices in $SU(3)$ flavor space with $\lambda^0 =\sqrt{2/3}I$; $G_S$ and $G_V$ are respectively the scalar-isoscalar and the vector-isoscalar coupling constant; $G_{IS}$ and $G_{IV}$ are respectively the scalar-isovector and the vector-isovector coupling constant. Since the Gell-Mann matrices with $a=1, 2, 3$ are identical to the Pauli matrices in $u$ and $d$ space, the isovector couplings break the $SU(3)$ symmetry while keeping the isospin symmetry. $K$ denotes the strength of the six-point KMT interaction \cite{KMT} that breaks the axial $U(1)_A$ symmetry, where `det' denotes the determinant in flavor space, i.e.,
\begin{eqnarray}
	\mathrm{det} (\bar{\psi} \Gamma \psi)
	&=&	\text{det}\begin{pmatrix}
		\bar{u}\Gamma u &\bar{u}\Gamma d &\bar{u}\Gamma s\\
		\bar{d}\Gamma u &\bar{d}\Gamma d &\bar{d}\Gamma s\\
		\bar{s}\Gamma u &\bar{s}\Gamma d &\bar{s}\Gamma s
		\end{pmatrix} \nonumber\\
	 &=& \sum_{i,j,k}\epsilon_\mathrm{ijk}(\bar{u}\Gamma q_i)(\bar{d}\Gamma q_j)(\bar{s}\Gamma q_k),
\end{eqnarray}
with $\Gamma = 1\pm \gamma_5$, and $\epsilon_\mathrm{ijk}$ being the Levi-Civita symbol with $q_i$, $q_j$, and $q_k$ representing the $u$, $d$, and $s$ quark fields. In the present study, we employ the parameters $m_u = m_d = 3.6$ MeV, $m_s = 87$ MeV, $G_S\Lambda^2 = 3.6$, $K\Lambda^5 = 8.9$, and the cutoff value in the momentum integral $\Lambda = 750$ MeV/c given in Refs.~\cite{Lut92,Bra13}, and define $R_{IS} = G_{IS}/G_S$ and $R_{IV} = G_{IV}/G_S$ as the reduced scalar-isovector and vector-isovector coupling constant, respectively.

\subsection{The Lagrangian density and the thermodynamic potential in the mean-field approximation}

To study the system at finite chemical potentials and temperature, we introduce the chemical potentials in the Lagrangian density
\begin{eqnarray} \label{Lfull}
	\mathcal{L} &=& \mathcal{L}_\mathrm{NJL} + \bar{\psi}\hat{\mu}\gamma_0\psi \nonumber\\
	&=&\bar{\psi}(i\gamma_\mu \partial^\mu + \hat{\mu}\gamma_0 - \hat{m})\psi \nonumber\\
	&& + \mathcal{L}_S + \mathcal{L}_V + \mathcal{L}_{\text{KMT}} + \mathcal{L}_{IS} + \mathcal{L}_{IV},
\end{eqnarray}
where $\hat{\mu} = \text{diag}(\mu_u, \mu_d, \mu_s)$ is the chemical potential matrix with
\begin{eqnarray}
	\mu_u &=& \frac{\mu_B}{3} + \frac{\mu_I}{2}, \nonumber\\
	\mu_d &=& \frac{\mu_B}{3} - \frac{\mu_I}{2}, \nonumber\\
	\mu_s &=& \frac{\mu_B}{3} - \mu_S,
\end{eqnarray}
or equivalently
\begin{eqnarray}
	\mu_B &=& \frac{3(\mu_u + \mu_d)}{2} , \nonumber\\
	\mu_I &=& \mu_u - \mu_d, \nonumber\\
	\mu_S &=& \frac{\mu_u + \mu_d}{2} - \mu_s,
\end{eqnarray}
where $\mu_B$, $\mu_I$, and $\mu_S$ are the baryon, isospin, and strangeness chemical potential, respectively.

Based on the mean-field approximation as detailed in Appendix \ref{MF Lagrangian}, the above Lagrangian density can be written as
\begin{equation}
	\mathcal{L}_\mathrm{MF}=\bar{\psi}\mathcal{S}^{-1}\psi -\mathcal{V},
\end{equation}
where
\begin{widetext}
\begin{eqnarray}\label{propagator}
	&&\mathcal{S}^{-1} (p) =\begin{pmatrix}
		\gamma_\mu p^\mu + \tilde{\mu}_u \gamma_0-M_u	 &	i \Delta \gamma_5 	& 0
		\\
		i \Delta\gamma_5 	&\gamma_\mu p^\mu +\tilde{\mu}_d \gamma_0 -M_d	&	0
		\\
		0& 0 &	\gamma_\mu p^\mu +\tilde{\mu}_s \gamma_0 -M_s
	\end{pmatrix}
\end{eqnarray}
\end{widetext}
is the inverse of the quark propagator $\mathcal{S}(p)$ as a function of quark momentum $p$, with
\begin{equation}\label{Delta}
\Delta = \left(G_{S} + 2G_{IS} -K \sigma_s\right)\pi
\end{equation}
being the gap parameter, and
\begin{eqnarray}
	\mathcal{V} &=&  G_{S} \left( \sigma_u^2 + \sigma_d^2  + \sigma_s^2 \right)  + \frac{G_{S}}{2}\pi^2 + G_{IS} (\sigma_u-\sigma_d)^2  \nonumber\\
	&+& G_{IS}\pi^2 - 4K \sigma_u \sigma_d \sigma_s - K\sigma_s \pi^2 \nonumber\\
	&-&\frac{1}{3}G_{V} \left( \rho_u+\rho_d+\rho_s \right)^2 -G_{IV}(\rho_{u}-\rho_{d})^2
\end{eqnarray}
being the condensation energy independent of the quark fields. In the above, $\rho_q=\langle \bar{q} \gamma_0 q\rangle$ and $\sigma_q=\langle \bar{q}q \rangle$ are the net-quark density and the chiral condensate, respectively, with $q=u,d,s$, and $\pi= \langle \bar{\psi} i \gamma_5 \lambda^1 \psi \rangle$ is the pion condensate. The Dirac effective mass or the constituent mass of quarks can be expressed as
\begin{eqnarray}
	M_u &=& m_u-2G_S\sigma_u - 2G_{IS}(\sigma_u-\sigma_d) + 2K\sigma_d \sigma_s ,\nonumber\\
	M_d &=& m_d-2G_S\sigma_d + 2G_{IS}(\sigma_u-\sigma_d) + 2K\sigma_u \sigma_s ,\nonumber\\
	M_s &=& m_s-2G_S\sigma_s + 2K\sigma_u \sigma_d + \frac{K}{2} \pi^2.\nonumber
\end{eqnarray}
Note that $M_u=M_d$ is used in some previous studies for the 2-flavor system~\cite{HeLY-ZhuangPF2005,Ava19} or at $\mu_B \sim 0$ \cite{Xia13,Lop21,Cao21}, while we consider the most general case in the present study on the QCD phase diagram at high baryon and isospin chemical potentials. The effective chemical potentials can be expressed as
\begin{eqnarray}
	\tilde{\mu}_u &=& \frac{\mu_B}{3} + \frac{\mu_I}{2} -\frac{2}{3}G_V  \rho-2\,G_{IV} (\rho_{u}-\rho_{d}),\nonumber\\
	\tilde{\mu}_d &=& \frac{\mu_B}{3} - \frac{\mu_I}{2} -\frac{2}{3}G_V  \rho + 2\,G_{IV}  (\rho_{u}-\rho_{d}),\nonumber\\
	\tilde{\mu}_s &=& \frac{\mu_B}{3} - \mu_S-\frac{2}{3}G_V\rho.\nonumber
\end{eqnarray}
Similarly, the effective baryon, isospin, and strangeness chemical potentials are
\begin{eqnarray}\label{eff}
	\tilde{\mu}_B &=& \frac{3(\tilde{\mu}_u + \tilde{\mu}_d)}{2} = \mu_B - 2G_V\rho, \nonumber\\
	\tilde{\mu}_I &=& \tilde{\mu}_u - \tilde{\mu}_d = \mu_I - 4G_{IV}\left(\rho_u-\rho_d\right) , \nonumber\\
	\tilde{\mu}_S &=& \frac{\tilde{\mu}_u + \tilde{\mu}_d}{2} - \tilde{\mu}_s = \mu_S.
\end{eqnarray}

Starting from the partition function as detailed in Appendix \ref{3-flavor NJL pion}, the thermodynamic potential can be expressed as
\begin{eqnarray}\label{OmegaNJL}
	\Omega &=& -\frac{1}{\beta V} \text{ln} \mathcal{Z} \nonumber\\
	&=&	\sum_{k=1}^4 \Omega_k  + \Omega_s
	+G_{S} ( \sigma_u^2 + \sigma_d^2  + \sigma_s^2 )  + \frac{G_{S}}{2}\pi^2 \nonumber\\
&+& G_{IS} (\sigma_u-\sigma_d)^2  + G_{IS}\pi^2- 4K \sigma_u \sigma_d \sigma_s \nonumber\\
&-& K \pi^2 \sigma_s -\frac{1}{3}G_{V} \left( \rho_u+\rho_d+\rho_s \right)^2 -G_{IV}(\rho_{u}-\rho_{d})^2,\nonumber\\
\end{eqnarray}
where $\beta=1/T$ is the inverse of the temperature, and
\begin{eqnarray}
	\Omega_k
	&=& -2N_c  \int \frac{d^3 p}{(2\pi)^3}\left[\frac{\lambda'_k}{2} + T \text{ln} \left( 1+e^{-\beta \lambda'_k} \right)\right], \nonumber\\
	\Omega_s
	&=& - 2N_c\,\int \frac{d^3 p}{(2\pi)^3} \left[ E_s + T \text{ln} \left( 1+e^{-\beta E_s^-} \right) \right.\nonumber\\
    &+& \left. T \text{ln} \left( 1+e^{-\beta E_s^+} \right)\right], \nonumber
\end{eqnarray}
are the kinetic contribution from light quarks and $s$ quarks, respectively. The quantities in the above are defined as $\lambda'_k = \lambda_k-\frac{\tilde{\mu}_B}{3}$ with $\lambda_k$ being the quasiparticle energy as detailed in Appendix \ref{3-flavor NJL pion}, and $E_s^{\pm}= E_s \pm \tilde{\mu}_s$ with $E_s=\sqrt{M_s^2 + \vec{p}^2}$ being the single $s$ quark energy. In the present study without considering the color superconductivity, the color degree of freedom contributes a factor of $N_c=3$.

\subsection{Gap equations}

Using the quark propagator as detailed in Appendix \ref{3-flavor NJL propagator pion}, the expressions of the chiral condensates and the net-quark densities for $u$, $d$, and $s$ quarks in terms of the phase-space distribution function can be written as
\begin{eqnarray}
	\sigma_{u}
	&=& 4\,  N_c\sum_{k=1}^4\int \frac{d^3 p}{(2\pi)^3}g_{\sigma u}\left(\lambda'_k\right) f\left(\lambda'_k\right),\label{gap_equation1}\\
	\sigma_{d}
	&=& 4\,  N_c\sum_{k=1}^4\int \frac{d^3 p}{(2\pi)^3}g_{\sigma d}\left(\lambda'_k\right) f\left(\lambda'_k\right),\\
	\sigma_{s}
	&=& 2\,N_c\int \frac{d^3 p}{(2\pi)^3}\frac{M_s}{E_s} \left[f\left(E_s^-\right)+f\left(E_s^+\right)-1\right],	\\
	\rho_{u}
	&=&4\,N_c\sum_{k=1}^4\int \frac{d^3 p}{(2\pi)^3}g_{\rho u}\left(\lambda'_k\right) \left[-\frac{1}{2}+f\left(\lambda'_k\right)\right],\\
	\rho_{d}
	&=&4\,N_c\sum_{k=1}^4\int \frac{d^3 p}{(2\pi)^3}g_{\rho d}\left(\lambda'_k\right) \left[-\frac{1}{2}+f\left(\lambda'_k\right)\right],\\
	\rho_{s}
	&=&2\,N_c\int \frac{d^3 p}{(2\pi)^3}\left[f\left(E_s^-\right)-f\left(E_s^+\right)\right].\label{gap_equation6}
\end{eqnarray}
The net baryon density $\rho_B$ and the isospin density $\rho_I$ can be calculated from
\begin{eqnarray}
\rho_B&=&(\rho_u+\rho_d+\rho_s)/3, \\
\rho_I&=&(\rho_u-\rho_d)/2.
\end{eqnarray}
The expression of the pion condensate $\pi$ is
\begin{equation}\label{pi}
	\pi
	=	4\,N_c \sum_{k=1}^4\int \frac{d^3 p}{(2\pi)^3}g_{\pi}\left(\lambda'_k\right) f\left(\lambda'_k\right).
\end{equation}
In the above expressions,
\begin{equation}
f(E)=\frac{1}{\exp(\beta E)+1}
\end{equation}
is the Fermi-Dirac distribution, the $g$ functions have the form of
\begin{eqnarray}
	g_{\sigma u}\left(\lambda'_k\right)
	&=& \frac{ \left[ \left(\lambda'_k+\tilde{\mu}_d\right)^2 - E_d^2 \right]M_u  -\Delta^2 M_d}
	{\prod_{j=1,j\neq k}^4 \left(\lambda'_k-\lambda'_j\right)},\\
	g_{\sigma d}\left(\lambda'_k\right)
	&=& \frac{ \left[ \left(\lambda'_k+\tilde{\mu}_u\right)^2 - E_u^2 \right]M_d  -\Delta^2 M_u}
	{\prod_{j=1,j\neq k}^4 \left(\lambda'_k-\lambda'_j\right)} ,\\
	g_{\pi}\left(\lambda'_k\right)
	&=& 2\frac{\left[ \vec{p}^2+M_uM_d-(\lambda'_k+\tilde{\mu}_u) (\lambda'_k+\tilde{\mu}_d)\right]\Delta  + \Delta^3 }
	{\prod_{j=1,j\neq k}^4 \left(\lambda'_k-\lambda'_j\right)},  \label{gpi} \nonumber\\ \\
%	g_{\Delta}\left(\lambda'_k\right)
%	&=& \frac{g_{\pi}\left(\lambda'_k\right)}{\pi} = 2G_{\pi}\frac{\left[ \vec{p}^2+M_uM_d-(\lambda'_k+\tilde{\mu}_u) (\lambda'_k+\tilde{\mu}_d)\right]  + \Delta^2 }
%	{\prod_{j=1,j\neq k}^4 \left(\lambda'_k-\lambda'_j\right)},    \nonumber\\
	g_{\rho u}\left(\lambda'_k\right)
	&=& \frac{ \left[ \left(\lambda'_k+\tilde{\mu}_d\right)^2 - E_d^2 \right]\left(\lambda'_k+\tilde{\mu}_u\right)  -\Delta^2 \left(\lambda'_k+\tilde{\mu}_d\right)}
	{\prod_{j=1,j\neq k}^4 \left(\lambda'_k-\lambda'_j\right)}	,\nonumber\\\\
	g_{\rho d}\left(\lambda'_k\right)
	&=& \frac{ \left[ \left(\lambda'_k+\tilde{\mu}_u\right)^2 - E_u^2 \right]\left(\lambda'_k+\tilde{\mu}_d\right)  -\Delta^2 \left(\lambda'_k+\tilde{\mu}_u\right)}
	{\prod_{j=1,j\neq k}^4 \left(\lambda'_k-\lambda'_j\right)} , \nonumber\\
\end{eqnarray}
and they satisfy the following relations
\begin{eqnarray}
	\sum_{k=1}^4g_{\sigma u}\left(\lambda'_k\right)=\sum_{k=1}^4g_{\sigma d}\left(\lambda'_k\right) =\sum_{k=1}^4g_{\pi}		\left(\lambda'_k\right)&=&0, \nonumber\\
	\sum_{k=1}^4g_{\rho u}	\left(\lambda'_k\right) =  \sum_{k=1}^4g_{\rho d}	\left(\lambda'_k\right) &=& 1,\nonumber\\
	g_{\rho u}\left(\lambda'_k\right) + g_{\rho d}\left(\lambda'_k\right)  &=& \frac{1}{2}.
\end{eqnarray}
Eqs.~(\ref{gap_equation1})-(\ref{gap_equation6}) can be obtained equivalently from
\begin{eqnarray}
\frac{\partial \Omega}{\partial \sigma_q} = \frac{\partial \Omega}{\partial \rho_q} = \frac{\partial \Omega}{\partial \pi} = 0,
\end{eqnarray}
with $q=u,d,s$ being the quark flavor, leading to the relations
\begin{eqnarray}
\sigma_q =\frac{\partial \Omega}{\partial M_q},~\rho_q = -\frac{\partial \Omega}{\partial \mu_q},~\pi=-\frac{\partial \Omega}{\partial \Delta}.
\end{eqnarray}
%Since $g_{\pi}$ is proportional to the pion condensate $\pi$ according to Eq.~(\ref{Delta}), $\pi=0$ is always a solution of Eq.~(\ref{pi}) at small $\mu_I$. With the increase of $\mu_I$, there appears nonzero solutions of $\pi$. With the further increase of $\mu_I$, we will see that $\pi$ becomes zero again.

\subsection{Equation of state}

The energy density can be obtained from the thermodynamic potential through the thermodynamical relation
\begin{eqnarray}
	\varepsilon &=& \Omega + \beta \frac{\partial}{\partial \beta}\Omega + \sum_i \mu_i \rho_i - \varepsilon_0 \nonumber\\
	&=& -2N_c \sum_{k=1}^4 \int \frac{d^3p}{(2\pi)^3}  \lambda'_k \left[  \frac{1}{2} - f\left(\lambda'_k\right)  \right] \nonumber\\
    &-&  2N_c\int \frac{d^3p}{(2\pi)^3} [E_s- E_s^- f(E_s^-) - E_s^+f(E_s^+) ] \nonumber\\
    &+& \sum_i \mu_i \rho_i + \mathcal{V} - \varepsilon_0.
\end{eqnarray}
where $\varepsilon_0$ is to ensure that the energy density is zero in vacuum. The pressure of the quark matter is
\begin{eqnarray}
	P =-\Omega + \Omega_0, \nonumber
\end{eqnarray}
where $\Omega_0 = \varepsilon_0$ is the thermodynamic potential in vacuum, to ensure that the pressure is zero in vacuum.

\section{Results and discussions}
\label{results}

Based on the theoretical framework of the 3-flavor NJL model, we discuss the behavior of the pion condensate in both baryon-free and baryon-rich quark matter as well as the corresponding 3-dimensional phase diagram. We neglect the vector-isoscalar interaction $G_V=0$, and the $s$ quark chemical potential is set to be $\mu_s=0$, throughout the study.

\subsection{Pion condensate in baryon-free quark matter}
\label{LQCD}

\begin{figure}[ht]
	\includegraphics[scale=0.3]{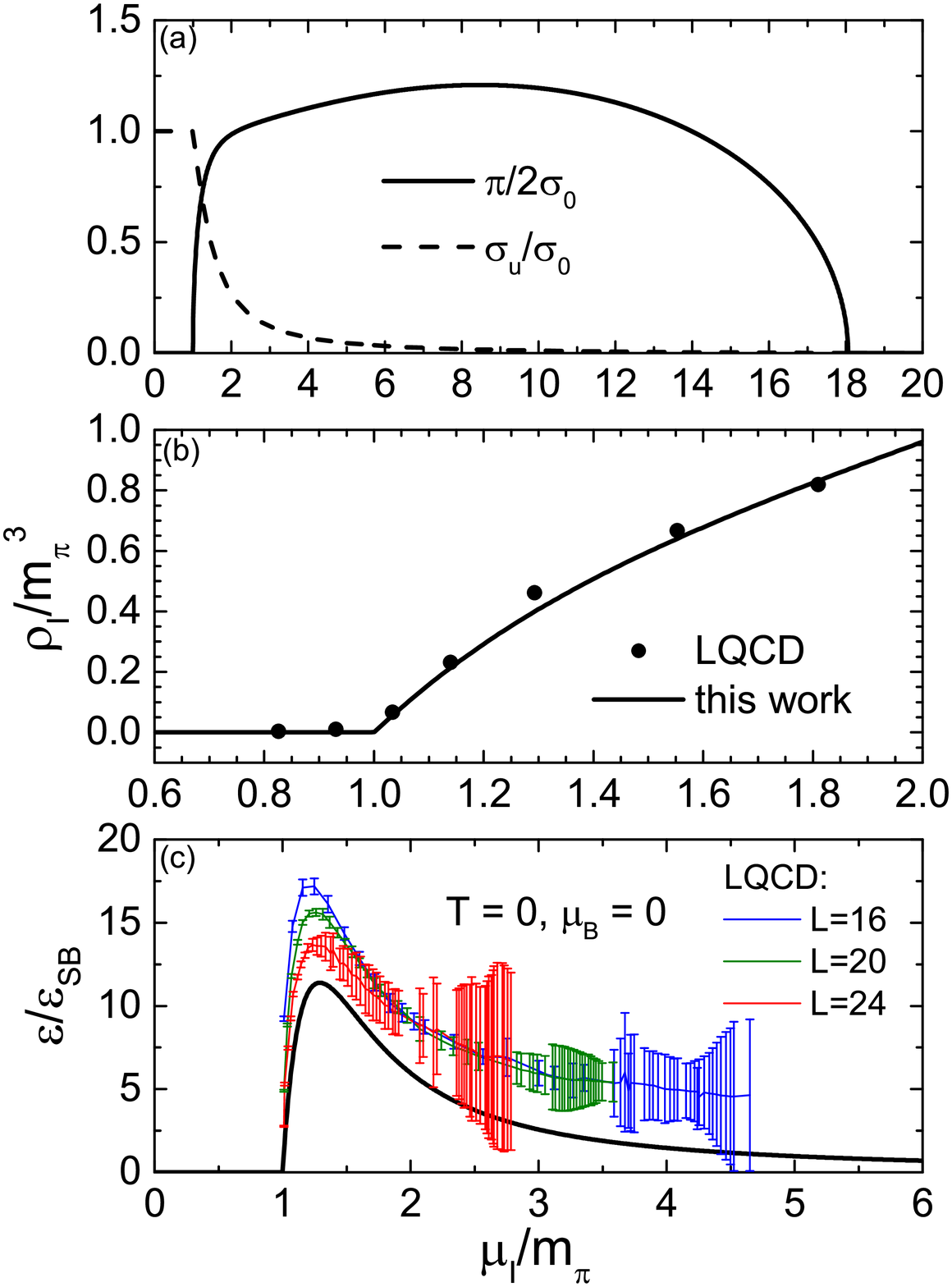}
	\caption{(Color online) Reduced pion and chiral condensate ($\pi/2\sigma_0$ and $\sigma_u/\sigma_0$) (a), reduced isospin density $\rho_I/m_\pi^3$ (b), and reduced energy density $\varepsilon/\varepsilon_{SB}$ (c) as a function of the reduced isospin chemical potential $\mu_I/m_\pi$ in cold ($T=0$) and baryon-free ($\mu_B=0$) quark matter. Available results from lattice QCD calculations are compared in panels (b) and (c). }\label{fig1}
\end{figure}

We start from fitting the physical pion mass and the lattice results by adjusting the isovector coupling constants. By choosing $R_{IS}=-0.002$, the pion mass is fitted to be $m_\pi=140.9$ MeV. Once $\mu_I$ is larger than this value, the reduced pion condensate $\pi/2\sigma_0$ becomes nonzero in cold and baryon-free quark matter, with $\sigma_0$ being the light quark condensate in vacuum, as shown in Fig.~\ref{fig1}(a). This behavior can be intuitively understood~\cite{He05} from the expression of the pion condensate $\pi$ [Eq.~(\ref{pi})] together with the relation Eqs.~(\ref{gpi}) and (\ref{Delta}). It can be seen that $\pi=0$ is always a solution of Eq.~(\ref{pi}), while for $\mu_I>m_\pi$ there appear nonzero solutions of $\pi$. For very large $\mu_I \sim 2$ GeV, the nonzero solutions of $\pi$ disappear again. Also shown in Fig.~\ref{fig1}(a) is the decrease of the reduced chiral condensate of $u$ quarks $\sigma_u/\sigma_0$ after the appearance of the pion condensate. This corresponds to a second-order phase transition, where the values of the chiral and pion condensate change continuously while their derivatives have a sudden jump as $\mu_I$ increases. For cold and baryon-free quark matter, the densities are zero below the threshold isospin chemical potential, and become finite above $\mu_I \sim m_\pi$ as a result of the nonzero pion condensate. By choosing $R_{IV}=0.25$, the reduced isospin density
as a function of the reduced isospin chemical potential reproduces very well the result from LQCD calculations~\cite{Bra18}, as shown in Fig.~\ref{fig1}(b). With the fitted $R_{IS}$ and $R_{IV}$, which are used throughout this study, the reduced energy density $\varepsilon/\varepsilon_{SB}$ as a function of the reduced isospin chemical potential is shown in Fig.~\ref{fig1}(c), with
\begin{equation}
\varepsilon_{SB}= \frac{N_f N_c}{4\pi^2} \left( \frac{\mu_I}{2}\right)^4
\end{equation}
being the energy density in the Stefan-Boltzmann limit. Comparing with the LQCD results in Fig.~22 of Ref.~\cite{Det12}, our result gives a similar peak position of $\mu_I/m_\pi \sim 1.3$, while the peak value of $\varepsilon/\varepsilon_{SB}$ decreases with increasing spatial extent $L$ from the LQCD calculations~\cite{Det12}.

\begin{figure}[ht]
	\includegraphics[scale=0.3]{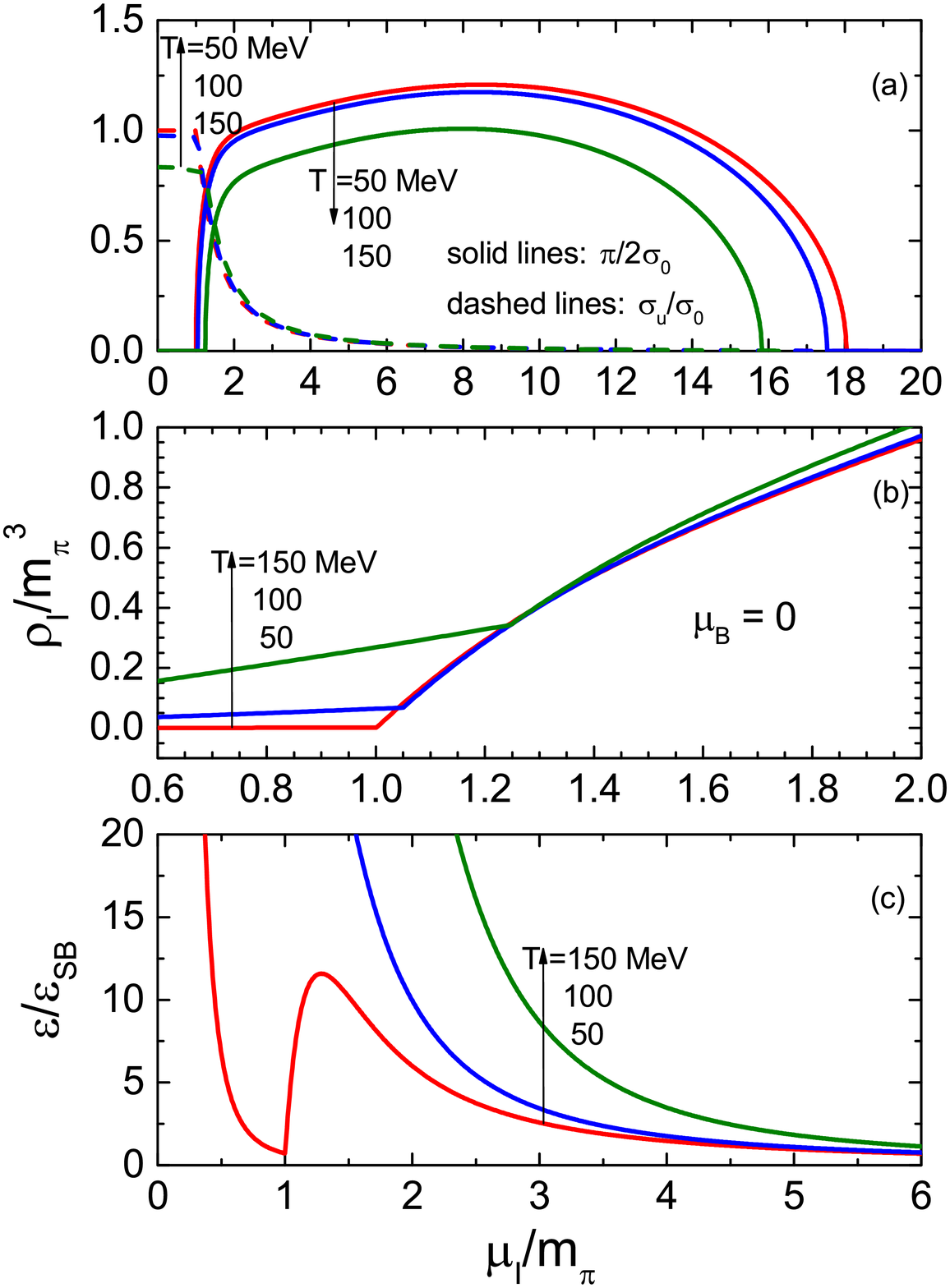}
	\caption{(Color online) Same as Fig.~\ref{fig1} but in hot ($T=50$, 100, and 150 MeV) and baryon-free quark matter.}\label{fig2}
\end{figure}

Similar results but in hot and baryon-free quark matter at $T=50$, 100, and 150 MeV are displayed in Fig.~\ref{fig2}. Due to the diffuseness of the Fermi-Dirac distribution at finite temperatures in Eqs.~(\ref{gap_equation1}) and (\ref{pi}), the reduced chiral condensate $\sigma_u/\sigma_0$ of $u$ quarks is smaller at small $\mu_I$, while the reduced pion condensate $\pi/2\sigma_0$ also decreases, compared to that at $T=0$. It is also interesting to see that the threshold isospin chemical potential $\mu_I$ increases with the increasing temperature. Since at finite temperatures the densities become nonzero even for small chemical potentials, the isospin density $\rho_I$ becomes finite and increases with the increasing temperature below the threshold isospin chemical potential. As a consequence, the energy density shows a similar behavior below the threshold isospin chemical potential. It is seen that there is still a peak around $\mu_I=1.3m_\pi$ at $T=50$ MeV, while such peak disappears at higher temperatures. It is worth noting that $\varepsilon_{SB}$ approaches $\mu_I \sim 0$ in the power of $\mu_I^4$. Although $\varepsilon$ is zero at $T=0$ below the threshold isospin chemical potential, it becomes finite at finite temperatures. This leads to the divergence behavior of $\varepsilon/\varepsilon_{SB}$ when $\mu_I$ approaches 0 as shown in Fig.~\ref{fig2}(c).

\begin{figure}[ht]
	\includegraphics[scale=0.3]{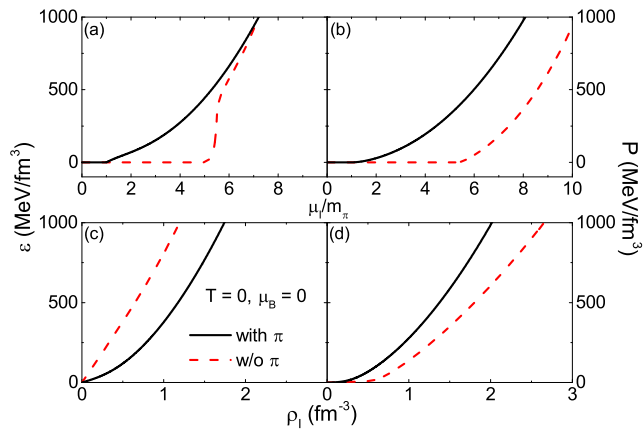}
	\caption{(Color online) Energy density $\varepsilon$ (left) and pressure $P$ (right) as a function of the reduced isospin chemical potential $\mu_I/m_\pi$ (upper) and the isospin density $\rho_I$ (lower) in cold and baryon-free quark matter with and without the pion condensate.}\label{fig3}
\end{figure}

The equation of state (EOS) of cold and baryon-free quark matter is displayed in Fig.~\ref{fig3}. It is seen that the energy density and the pressure become finite at a much larger isospin chemical potential $\mu_I \sim 2 M$, with $M$ being the Dirac mass of light quarks in vacuum, in the absence of the pion condensate, compared to the case with the pion condensate incorporated. This leads to a larger energy density and pressure for a given isospin chemical potential in the presence of the pion condensate. As for the EOS as a function of the isospin density $\rho_I$, the energy density is reduced while the pressure is increased with the pion condensate, compared to the case without considering the pion condensate. This is due to the different relations between $\rho_I$ and $\mu_I$ in the two cases. It can be seen from Fig.~\ref{fig3} the pion condensate will stiffen the $P \sim \varepsilon$ relation. It could thus be further speculated that this might affect the properties of compact stars from solving the Tolman-Oppenheimer-Volkoff equation. However, this generally does not happen~\cite{Ebe06,Ade10}, since it requires a large isospin chemical potential but a not too large baryon chemical potential reached at the same time, related to the discussions in the next subsection.

\subsection{Pion condensate in baryon-rich quark matter}
\label{3DQCD}

\begin{figure}[ht]
	\includegraphics[scale=0.35]{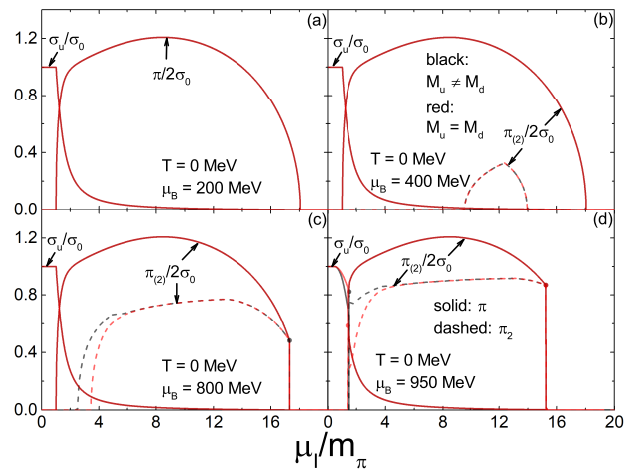}
	\caption{(Color online) Reduced pion and chiral condensate ($\pi/2\sigma_0$ and $\sigma_u/\sigma_0$) as a function of the reduced isospin chemical potential $\mu_I/m_\pi$ in cold ($T=0$ MeV) and baryon-rich ($\mu_B=200$ (a), 400 (b), 800 (c), and 950 (d) MeV) quark matter. Results are compared with those obtained using the assumption $M_u=M_d$.}\label{fig4}
\end{figure}

\begin{figure}[ht]
	\includegraphics[scale=0.35]{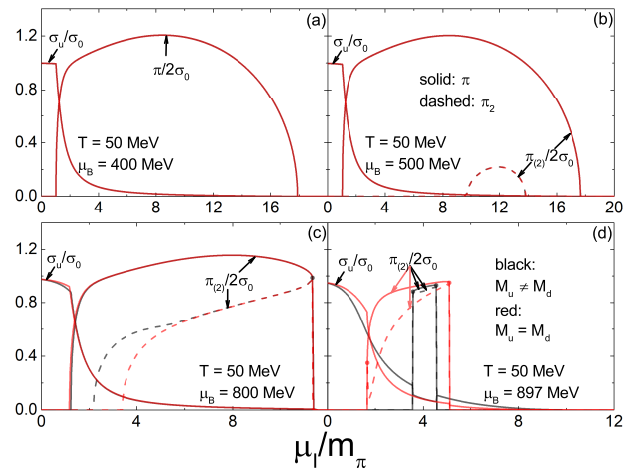}
	\caption{(Color online) Similar to Fig.~\ref{fig4} but in hot ($T=50$ MeV) and baryon-rich ($\mu_B=400$ (a), 500 (b), 800 (c), and 897 (d) MeV) quark matter. Results are compared with those obtained using the assumption $M_u=M_d$.}\label{fig5}
\end{figure}

For the ease of discussions on the complete 3-dimensional QCD phase diagram, we display in Figs.~\ref{fig4} and \ref{fig5} the reduced pion and chiral condensate in cold ($T=0$ MeV) and hot ($T=50$) and baryon-rich quark matter. Our rigourous calculation reduces to those in other studies using the assumption $M_u=M_d$ at $\mu_B=0$, but some differences are expected to appear at large baryon chemical potentials, and results from the two cases are compared in this and next subsections. It is seen from Fig.~\ref{fig4}(a) and Fig.~\ref{fig5}(a) that the results for smaller $\mu_B$ are not qualitatively different from those in baryon-free quark matter shown in Fig.~\ref{fig1} (a) and Fig.~\ref{fig2}(a). As $\mu_B$ increases, except that the pion condensate becomes zero again at a slightly smaller $\mu_I$, there appears a second nonzero solution ($\pi_2$) of the pion condensate from Eq.~(\ref{pi}) between about $\mu_I \sim 10m_\pi$ and $\mu_I \sim 14 m_\pi$, as shown in Fig.~\ref{fig4}(b) and Fig.~\ref{fig5}(b). At an even higher $\mu_B$, the occurrence of $\pi_2$ is at an even lower $\mu_I$. On the other hand, there is a first-order phase transition of the pion condensate at large $\mu_I$ for both $\pi$ and $\pi_2$, where the pion condensate changes suddenly from a finite value to zero, as shown by the vertical lines in Fig.~\ref{fig4}(c) and Fig.~\ref{fig5}(c). In addition, as shown in the same panel, it is seen that the assumption of $M_u=M_d$ overestimates significantly the threshold $\mu_I$ for $\pi_2$, and underestimates the threshold $\mu_I$ for $\pi$, especially at $T=50$ MeV compared with those at $T=0$ MeV. If the baryon chemical potential is further increased, the occurrence of the pion condensate becomes a first-order phase transition as well, and this is displayed in Fig.~\ref{fig4}(d) and Fig.~\ref{fig5}(d). It is also seen in Fig.~\ref{fig5}(d) that the region of $\pi \neq 0$ and  $\pi_2 \neq 0$ is much smaller from the rigourous calculations of $M_u \neq M_d$ compared to those by assuming $M_u=M_d$. In addition, the chiral condensate generally has a sudden increase (decrease) once the pion condensate has a sudden decrease (increase) with the increasing isospin chemical potential. Figure~\ref{fig5}(d) shows that the QCD phase structure can be different at large $\mu_B$ and higher temperatures from the rigourous calculations comparing with those by assuming equal constituent mass for $u$ and $d$ quarks.

\begin{figure}[ht]
	\includegraphics[scale=0.35]{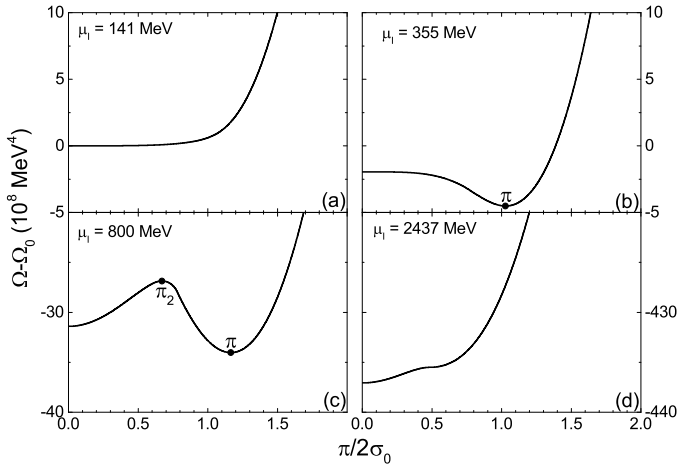}
	\caption{Thermodynamic potential $\Omega$ (subtracting the contribution $\Omega_0$ in vacuum) as a function of the reduced pion condensate at different isospin chemical potentials $\mu_I$ in cold ($T=0$ MeV) and baryon-rich ($\mu_B=800$ MeV) quark matter, corresponding to the condition in Fig.~\ref{fig4}(c).}\label{fig6}
\end{figure}

Although $\pi$ and $\pi_2$ calculated based on Eq.~(\ref{pi}) satisfy the condition $\partial \Omega / \partial \pi=0$, they are not both stable. For the $T$, $\mu_B$, and 4 typical isospin chemical potential $\mu_I$ chosen according to Fig.~\ref{fig4}(c), we show in Fig.~\ref{fig6} the thermodynamic potential $\Omega$ as a function of the reduced pion condensate, after subtracting the contribution $\Omega_0$ in vacuum. At $\mu_I=141$ MeV, it is seen that a local minimum point of $\Omega$ is about to appear, leading to the occurrence of $\pi$. At $\mu_I=355$ MeV, except for a local minimum point corresponding to $\pi$, a local maximum point of $\Omega$ is about to appear, leading to the occurrence of $\pi_2$. At $\mu_I=800$ MeV, there are both local maximum and local minimum points of $\Omega$, showing the existence of both $\pi$ and $\pi_2$ states. At $\mu_I=2437$ MeV, both the local maximum and local minimum of $\Omega$ are about to disappears, and $\pi$ and $\pi_2$ turn to 0 accordingly. Since the solution $\pi_2$ corresponds to a maximum thermodynamic potential, it is an unstable solution, and the system favors $\pi$ rather than $\pi_2$. It is also argued that the instability of $\pi_2$ could be cured by considering the free energy of a system with a fixed baryon density~\cite{He06} or in a Fermi system with a finite-range momentum-dependent interaction~\cite{For05}.

\begin{figure}[ht]
	\includegraphics[scale=0.35]{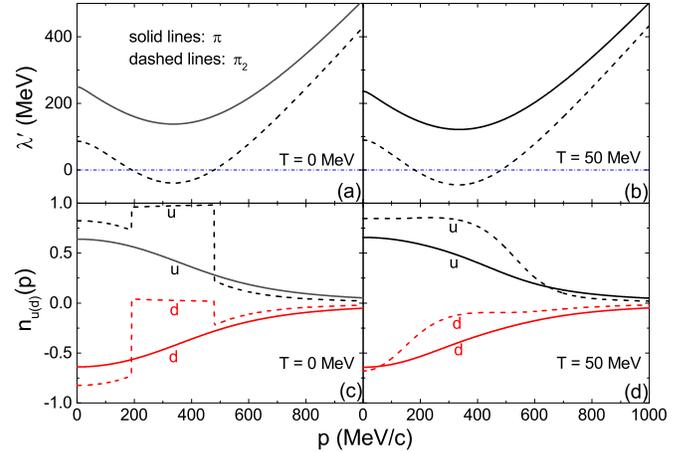}
	\caption{(Color online) Dispersion relations (upper) and net-quark momentum distributions (lower) for $u$ and $d$ quarks in baryon-rich ($\mu_B=800$ MeV) and isospin-asymmetric ($\mu_I=800$ MeV) quark matter at $T=0$ [(a), (c)] and 50 MeV [(b), (d)].}\label{fig7}
\end{figure}

To understand in more details the properties of $\pi$ and $\pi_2$, we display in Fig.~\ref{fig7} the dispersion relations and net-quark momentum distributions for $u$ and $d$ quarks in the condition of Fig.~\ref{fig6}(c) and also at $T=50$ MeV. For the dispersion relation, we show one of the 4 solutions of $\lambda'_k = \lambda_k-\frac{\tilde{\mu}_B}{3}$ with $k=1,2,3,4$ and $\lambda_k$ being the quasiparticle energy as detailed in Appendix \ref{3-flavor NJL pion}. The net-quark momentum distribution is defined as
\begin{equation}
	n_{u(d)}
	=2\sum_{k=1}^4 g_{\rho u(d)}\left(\lambda'_k\right) \left[-\frac{1}{2}+f\left(\lambda'_k\right)\right],
\end{equation}
so integrating $n_{u(d)}$ gives the net-quark density, i.e.,
\begin{equation}
\rho_{u(d)} = 2 N_c \int \frac{d^3 p}{(2\pi)^3} n_{u(d)}.
\end{equation}
It is seen from Fig.~\ref{fig7}(a) that the quasiparticle energy for $\pi_2$ becomes negative when the momentum $p$ is between about 200 and 500 MeV/c. For the other 3 quasiparticle energy solutions for $\pi_2$, they do not change sign as a function of the momentum. The negative quasiparticle energy state corresponds to the so-called Sarma phase~\cite{sar63}, where the quasiparticle excitation does not need additional energy. The Sarma phase breaks the pairing of $u$ and $\bar{d}$ quarks, as can be see from Fig.~\ref{fig7}(c) that $n_d$ and $n_u$ is approximately of a constant value 0 and 1 in the region of negative $\lambda'$, respectively, compared with the pairing states when $n_u(p) = -n_d(p)$ is always satisfied. At $T=50$ MeV, the dispersion relation of $\lambda'$ is similar to that at $T=0$ MeV, while there is no sudden jump in the net-quark momentum distribution, though $n_u(p)$ and $n_d(p)$ become asymmetric around the momentum region where $\lambda'$ is negative. This behavior corresponds to the 'quasi' Sarma phase at finite temperatures~\cite{Mu09,Boe15}. Also shown are the dispersion relation of $\lambda'$ for the $\pi$ solution corresponding to the local minimum of the thermodynamic potential as shown in Fig.~\ref{fig6}. In such case, it is seen that $\lambda'$ does not change sign, and $n_u(p) = -n_d(p)$ is always satisfied at $T=0$ MeV and approximately satisfied at $T=50$ MeV.

\subsection{3-dimensional phase diagram}

\begin{figure}[ht]
	\includegraphics[scale=0.35]{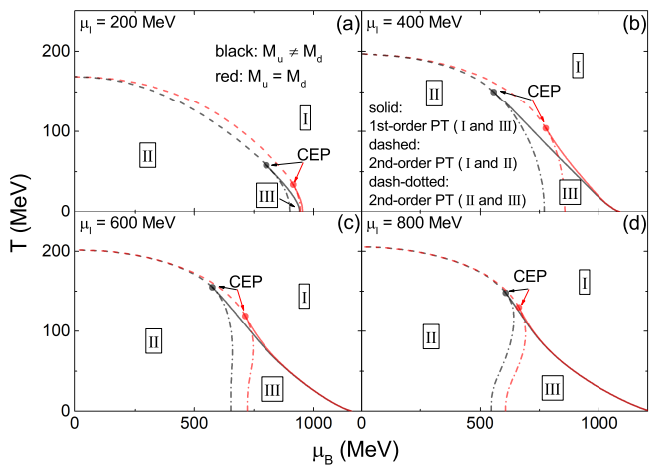}
	\caption{(Color online) Phase diagrams in the $T-\mu_B$ plane at different isospin chemical potentials $\mu_I=200$ (a), 400 (b), 600 (c), and 800 (d) MeV, from rigourous calculations of $M_u \neq M_d$ and approximated calculations of $M_u=M_d$. Solid lines represent the first-order phase transition (PT) between Phase I and Phase III, dashed lines represent the second-order phase transition between Phase I and Phase II, and dash-dotted lines represent the second-order phase transition between Phase II and Phase III.}\label{fig8}
\end{figure}

We then move to the discussions of the 3-dimensional QCD phase diagram. Since the appearance of the pion condensate always leads to the decrease of the chiral condensate, here we only discuss the behavior of the pion condensate in the $T-\mu_B-\mu_I$ space, where the chiral condensate generally has an opposite behavior. Figure~\ref{fig8} displays the phase diagrams in the $T-\mu_B$ plane at different isospin chemical potentials, where we show 3 phases as discussed above. Phase I is the normal baryon-rich and isospin-asymmetric quark matter with $\pi=0$. Phase II is the pion superfluid phase with $\pi\ne0$. Phase III is the phase with both nonzero solutions of $\pi$ and $\pi_2$, with the latter corresponding to the existence of the Sarma phase as discussed above. It is seen that Phase I generally exists at larger $T$ or larger $\mu_B$, while Phase II generally exists at smaller $T$ and $\mu_B$.  Phase III exists in the area between the solid line and the dash-dotted line. It is seen that the phase transition between Phase I and Phase II, in the absence of Phase III, is always a second-order one, with the phase boundary represented by the dashed lines, so are the phase transition between Phase II and Phase III, with the phase boundary represented by the dash-dotted lines. On the other hand, the phase transition between Phase I and Phase III is always a first-order one, with the phase boundary represented by the solid lines. The critical end point (CEP), which connects the boundaries of the first-order phase transition and the second-order phase transitions, moves to a higher temperature with $\mu_I$ changing from 200 MeV to 400 MeV, and the increasing trend saturates above $\mu_I=400$ MeV. The assumption of $M_u=M_d$ leads to a different phase structure at larger $\mu_B$, resulting in a CEP at lower temperatures and larger baryon chemical potentials.

\begin{figure}[ht]
	\includegraphics[scale=0.35]{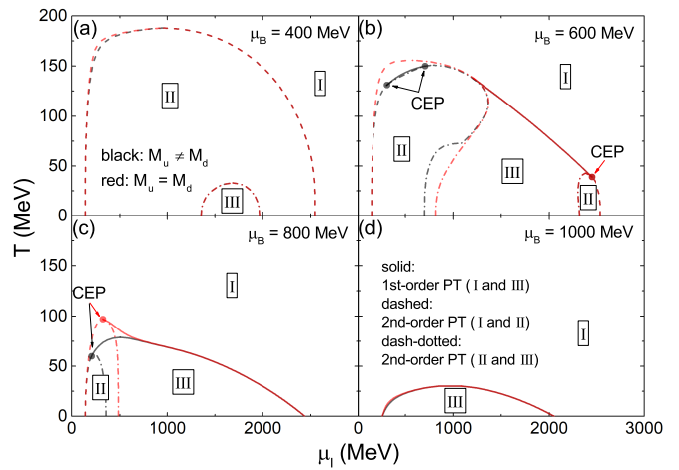}
	\caption{(Color online) Similar to Fig.~\ref{fig8} but in the $T-\mu_I$ plane at different baryon chemical potentials $\mu_B=400$ (a), 600 (b), 800 (c), and 1000 (d) MeV.}\label{fig9}
\end{figure}

Figure~\ref{fig9} displays the phase diagrams in the $T-\mu_I$ plane at different baryon chemical potentials. It is seen that the normal quark phase (Phase I) generally exists at very small or large isospin chemical potentials, or at high temperatures, while the area of the pion superfluid phase (Phase II) shrinks dramatically with the increasing baryon chemical potential. The phase transitions are always of second-order at smaller baryon chemical potentials, while the first-order phase transition becomes more and more dominate with the increasing baryon chemical potential. Phase III with $\pi_2 \ne 0$ doesn't exist at $\mu_B=0$ (not shown here), but it gradually appears inside Phase II at small baryon chemical potentials, and its area becomes larger and dominate at large baryon chemical potentials. The difference between results from rigourous calculations ($M_u \neq M_d$) and approximations ($M_u=M_d$) on the phase diagram is seen in panels (b) and (c), mainly exists in the relative area of Phase II and Phase III as well as the position of the CEP.

\begin{figure}[ht]
	\includegraphics[scale=0.35]{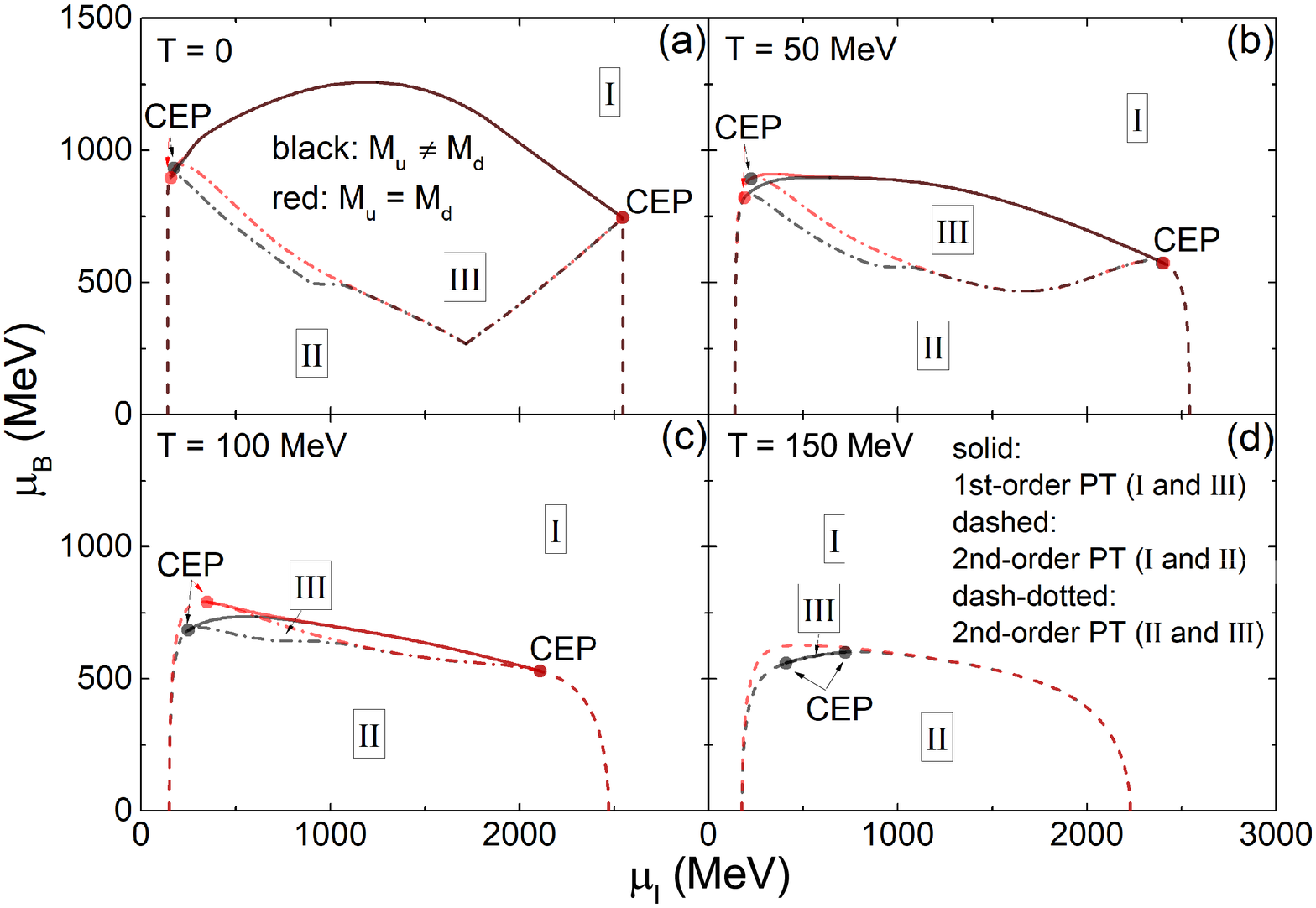}
	\caption{(Color online) Similar to Fig.~\ref{fig8} but in the $\mu_B-\mu_I$ plane at different temperatures $T=0$ (a), 50 (b), 100 (c), and 150 (d) MeV.}\label{fig10}
\end{figure}

Figure~\ref{fig10} displays the phase diagrams in the $\mu_B-\mu_I$ plane at different temperatures. Again, the normal quark phase (Phase I) exists at larger $\mu_B$ and/or very small or large $\mu_I$, and the pion superfluid phase (Phase II) is observed at smaller $\mu_B$ and intermediate $\mu_I$, already seen in Figs.~\ref{fig8} and \ref{fig9}. The area of Phase II shrinks with the increasing temperature. Also, the first-order phase transition and Phase III become less dominate at higher temperatures. The deviations of results using the approximation ($M_u=M_d$) from rigourous calculations ($M_u \neq M_d$) can now be quantitatively seen within $\mu_B \in (500, 900)$ MeV. Figure~\ref{fig10} is of course consistent with Figs.~\ref{fig8} and \ref{fig9}, and they together give a whole picture of the 3-dimensional QCD phase diagram.

\section{Conclusion}
\label{summary}

With the scalar-isovector and vector-isovector coupling constants adjusted to fit the physical pion mass and the lattice QCD results in baryon-free quark matter, we have studied the 3-dimensional QCD phase diagram by considering the pion condensate based on the 3-flavor NJL model. We found that the pion condensate becomes less important at higher temperatures or larger baryon chemical potentials. Thus, although incorporating the pion condensate would stiffen the equation of state of strange quark matter, it generally does not affect the properties of compact star systems where large baryon chemical potentials are reached. Besides the normal solution, we observe the appearance of a second nonzero solution of the pion condensate with the increase of the isospin chemical potential in baryon-rich quark matter, while both solutions disappear at very large isospin chemical potentials. The normal solution corresponds to the local minimum of the thermodynamic potential and represents the pion superfluid phase, while the second solution corresponds to the local maximum of the thermodynamic potential and is related to the Sarma phase. The occurrence or the disappearance of the pion condensate is a second-order phase transition at higher temperatures or smaller baryon chemical potentials, while it becomes a first-order one at lower temperatures or larger baryon chemical potentials. The calculations by assuming equal constituent mass of $u$ and $d$ quarks may introduce large errors in the 3-dimensional QCD phase diagram within $\mu_B \in (500, 900)$ MeV, and affect the extraction of the critical end point, compared with the rigourous calculations in the present study.

To further explore the QCD phase structure, one can incorporate the polyakov loop into the NJL model, and study the interplay among the chiral condensate, the pion condensate, and the polyakov loop~\cite{Zha07b}. The kaon condensate can be further incorporated by considering systems at large strangeness chemical potentials. In addition, although the pion condensate generally will not affect the equation of state of strange quark matter and thus properties of compact stars, it is of great interest to further incorporate the chiral imbalance~\cite{Khu20}, the color superconductivity~\cite{Sun07}, etc., and see their effects on the QCD phase diagram and compact star properties. Further detailed properties of the QCD phase structure, e.g., the Larkin-Ovchinnikov-Fudde-Ferrell phase, are also worth investigating, as shown in Refs.~\cite{He06,Mu10}.

\appendix

\section{The Lagrangian from mean-field approximation}\label{MF Lagrangian}

In the mean-field approximation, it is assumed that deviations due to fluctuations of all quantities $A$ from their thermal average values $\langle A\rangle$ are small. Thus, the following relations can be introduced to linearize the Lagrangian
\begin{eqnarray}\label{MF}
	(\bar{\psi}\Gamma_i \psi) & \approx & \langle\bar{\psi} \Gamma_i \psi\rangle, \nonumber\\
	(\bar{\psi}\Gamma_i \psi)^2 & \approx &  2 \bar{\psi} \Gamma_i \psi \langle\bar{\psi} \Gamma_i \psi\rangle - \langle\bar{\psi} \Gamma_i \psi\rangle^2,  \nonumber \\
	(\bar{\psi} \Gamma_i \psi \bar{\psi} \Gamma_j \psi) & \approx & \bar{\psi} \Gamma_i \psi \langle\bar{\psi} \Gamma_j \psi\rangle + \bar{\psi} \Gamma_j \psi \langle\bar{\psi} \Gamma_i \psi\rangle \nonumber\\
	&& - \langle\bar{\psi} \Gamma_i \psi\rangle\langle\bar{\psi} \Gamma_j \psi\rangle, \nonumber\\
	(\bar{\psi} \Gamma_i \psi \bar{\psi} \Gamma_j \psi  \bar{\psi} \Gamma_k \psi) & \approx &(\bar{\psi} \Gamma_i \psi )\langle\bar{\psi} \Gamma_j \psi\rangle \langle\bar{\psi} \Gamma_k \psi\rangle \nonumber\\
	&& + (\bar{\psi} \Gamma_j \psi )\langle\bar{\psi} \Gamma_i \psi\rangle \langle\bar{\psi} \Gamma_k \psi\rangle     \nonumber \\
	&& + (\bar{\psi} \Gamma_k \psi )\langle\bar{\psi} \Gamma_i \psi\rangle \langle\bar{\psi} \Gamma_j \psi\rangle     \nonumber \\
	&& -2 \langle\bar{\psi} \Gamma_i \psi\rangle \langle\bar{\psi} \Gamma_j \psi\rangle \langle\bar{\psi} \Gamma_k \psi\rangle,
%	(\bar{\psi} \Gamma_i \psi \bar{\psi} \Gamma_j \psi)^2 & \approx &  \langle\bar{\psi}\Gamma_i \psi\rangle^2 (2\bar{\psi}\Gamma_j \psi \langle\bar{\psi}\Gamma_j\psi\rangle)     \nonumber \\
%	&&+\langle\bar{\psi}\Gamma_j \psi\rangle^2 (2\bar{\psi}\Gamma_i \psi \langle\bar{\psi}\Gamma_i\psi\rangle) \nonumber \\
%	&&- 3\langle\bar{\psi}\Gamma_i \psi\rangle^2 \langle\bar{\psi}\Gamma_j \psi\rangle^2,
\end{eqnarray}
with $\Gamma = \{1,\gamma_5,\gamma_\mu,\gamma_5\gamma_\mu\}$, and the angular bracket
	denoting the expectation value from the quantum statistical average. In our previous studies~\cite{Liu16}, we assumed $\langle\bar{\psi}\gamma^k
	\psi\rangle =\langle\bar{\psi}\gamma_5  \vec{\tau}\psi\rangle=\langle\bar{\psi}\gamma_5  \lambda^a\psi\rangle=
	\langle\bar\psi\gamma_5\gamma^\mu\psi\rangle=0$ due to the parity symmetry in a static quark matter, so the condensates $\langle \bar{\psi}_i \psi_j \rangle $ with $i \neq j$ vanishes since it is assumed that the flavor is conserved in the case of $\mu_I<m_\pi$. In order to study the 3-dimensional QCD phase diagram, we consider systems at larger isospin chemical potentials where pion condensates may arise, i.e.,
\begin{eqnarray}
	\pi^+ &=& \langle \bar{\psi} i \gamma_5 \lambda^1_+ \psi \rangle = \sqrt{2}\langle \bar{u} i \gamma_5 d \rangle = \frac{\pi}{\sqrt{2}} e^{i\theta_{ud}}, \nonumber\\
	\pi^- &=& \langle \bar{\psi} i \gamma_5 \lambda^1_- \psi \rangle = \sqrt{2}\langle \bar{d} i \gamma_5 u \rangle = \frac{\pi}{\sqrt{2}} e^{-i\theta_{ud}}, \nonumber\\
	\pi &=& \langle \bar{\psi} i \gamma_5 \lambda^1 \psi \rangle = \langle \bar{u} i \gamma_5 d \rangle + \langle \bar{d} i \gamma_5 u \rangle,  \nonumber
\end{eqnarray}
with $\lambda^1_{\pm} = \frac{1}{\sqrt{2}}(\lambda^1\pm i\lambda^2)$. In such case, the nonzero expectation value of $\langle \bar{u} i \gamma_5 d \rangle$ or $\langle \bar{d} i \gamma_5 u \rangle$ spontaneously break the $U_I(1)$ symmetry, corresponding to the Bose-Einstein condensation of charged pions. The phase $\theta_{ud}$ represents the direction of the $U_I(1)$ symmetry breaking. Since the thermodynamic potential does not depend on $\theta_{ud}$ but depends only on $|\pi^{\pm}|^2$ or $|\pi|^2$, we can set them to be real values corresponding to $\theta_{ud}=0$ without losing generality. The Kaon condensate, which could be important at large strangeness chemical potentials, is not considered in the present study.

In the mean-field approximation by using the relations of Eq.~(\ref{MF}), the scalar-isoscalar term can be expressed as
	\begin{equation} \label{}
		\frac{G_{S}}{2}\sum_{a=0}^8 [(\bar{\psi}\lambda^a\psi)^2 + (\bar{\psi}i \gamma_5 \lambda^a \psi)^2] = \bar{\psi}\Sigma_S \psi	- \mathcal{V}_S,
	\end{equation}
	where $\mathcal{V}_S = G_{S} \left( \sigma_u^2 + \sigma_d^2  + \sigma_s^2 \right)  + \frac{G_{S}}{2}\pi^2 $ is the scalar-isoscalar condensate energy, with
\begin{eqnarray}
\sigma_u &=& \langle \bar{u} u \rangle, \nonumber\\
\sigma_d &=& \langle \bar{d} d \rangle, \nonumber\\
\sigma_s &=& \langle \bar{s} s \rangle
\end{eqnarray}
being the chiral condensates for $u$, $d$, and $s$ quarks, respectively, and
	\begin{eqnarray}
		\Sigma_S =
		\begin{pmatrix}
			2G_{S}\sigma_u & i  G_{S} \pi \gamma_5 & 0
			\\
			i  G_{S}\pi\gamma_5 & 2G_{S}\sigma_d & 0
			\\
			0 & 0 & 2G_{S}\sigma_s
		\end{pmatrix}
	\end{eqnarray}
is the self-energy contributed from the scalar-isoscalar interaction. The scalar-isovector term in the mean-field approximation can be expressed as
	\begin{equation} \label{}
		G_{IS} \sum_{a=1}^3[(\bar{\psi}\lambda^a\psi)^2 + (\bar{\psi}i \gamma_5 \lambda^a \psi)^2] = \bar{\psi}\Sigma_{IS} \psi	- \mathcal{V}_{IS},
	\end{equation}
where $\mathcal{V}_{IS} = G_{IS} (\sigma_u-\sigma_d)^2  + G_{IS}\pi^2$ is the scalar-isovector condensate energy, and
	\begin{eqnarray}
		\Sigma_{IS} =
		\begin{pmatrix}
			2G_{IS}(\sigma_u-\sigma_d) & i  2G_{IS} \pi \gamma_5 & 0
			\\
			i  2G_{IS}\pi\gamma_5 & -2G_{IS}(\sigma_u-\sigma_d) & 0
			\\
			0& 0 & 0
		\end{pmatrix}
	\end{eqnarray}
is the self-energy contributed from the scalar-isovector interaction. The KMT term in the mean-field approximation can be expressed as
	\begin{equation}
		-K\big\{\mathrm{det} (\bar{\psi} (1 + \gamma_5) \psi) + \mathrm{det} (\bar{\psi} (1 - \gamma_5) \psi)\big\} =\bar{\psi} \Sigma_K \psi - \mathcal{V}_K,
	\end{equation}
where $\mathcal{V}_{K} = - 4K \sigma_u \sigma_d \sigma_s - K  \pi^2 \sigma_s  $ is the condensate energy contribution from the KMT interaction, and
	\begin{eqnarray}
		\Sigma_{K} =
		\begin{pmatrix}
			-2K\sigma_d \sigma_s & -iK\pi \sigma_s\gamma_5 &0
			\\
			-iK\pi \sigma_s\gamma_5 &  -2K\sigma_u \sigma_s   & 0
			\\
			0 & 0 & -2K\sigma_u \sigma_d - \frac{K}{2} \pi^2
		\end{pmatrix}
	\end{eqnarray}
is the self-energy contributed from the KMT interaction. Considering only the flavor-singlet state, the vector-isoscalar term in the mean-field approximation can be expressed as
	\begin{equation} \label{}
		-\frac{G_{V}}{2}\sum_{a=0}^8 [(\bar{\psi}\gamma^\mu \lambda^a\psi)^2 + (\bar{\psi}i \gamma_5 \gamma^\mu\lambda^a \psi)^2] = \bar{\psi} \tilde{\mu}_V\gamma_0 \psi - \mathcal{V}_V,
	\end{equation}
where $\mathcal{V}_V = -\frac{1}{3}G_{V} \left( \rho_u+\rho_d+\rho_s \right)^2 $ is the vector-isoscalar condensate energy, with $\rho_q$ ($q=u,d,s$) being the net-quark density, and
	\begin{eqnarray}
		\tilde{\mu}_{V} =
		\begin{pmatrix}
			-\frac{2}{3}G_V  \rho & 0 & 0
			\\
			0 & -\frac{2}{3}G_V  \rho & 0
			\\
			0& 0 & -\frac{2}{3}G_V  \rho
		\end{pmatrix}
	\end{eqnarray}
represents the contribution of the effective chemical potential from the vector-isoscalar interaction, with $\rho=\rho_u+\rho_d+\rho_s$ being the total net-quark density. The vector-isovector term in the mean-field approximation can be expressed as
	\begin{equation} \label{}
		-G_{IV} \sum_{a=1}^3[(\bar{\psi}\gamma^\mu\lambda^a\psi)^2 + (\bar{\psi} i\gamma_5 \gamma^\mu\lambda^a \psi)^2] = \bar{\psi} \tilde{\mu}_{IV}\gamma_0 \psi - \mathcal{V}_{IV},
	\end{equation}
	where $\mathcal{V}_{IV} = -G_{IV}(\rho_{u}-\rho_{d})^2 $ is the vector-isovector condensate energy, and
	\begin{eqnarray}
		\tilde{\mu}_{IV} =
		\begin{pmatrix}
			-2\,G_{IV}(\rho_{u}-\rho_{d}) & 0 & 0
			\\
			0 & 2\,G_{IV}(\rho_{u}-\rho_{d}) & 0
			\\ 0 & 0 & 0
		\end{pmatrix}
	\end{eqnarray}
	represents the contribution of the effective chemical potential from the vector-isovector interaction.

\section{Thermodynamic potential}\label{3-flavor NJL pion}

In this appendix, we obtain the thermodynamic potential from the partition function, which in the grand canonical ensemble is written as
\begin{equation}
	\mathcal{Z}=\text{Tr} e^{-\beta(\hat{H}-\mu \hat{N})} = \sum_a \int d\Psi_a \langle \Psi_a |  e^{-\beta(\hat{H}-\mu \hat{N})} | \Psi_a\rangle,
\end{equation}
where $\beta=T^{-1}$ is the inverse of the temperature, $\mu$ is the chemical potential, and $\hat{H}$ and $\hat{N}$ are the Hamiltonian operator and the quark number operator, respectively. The sum $\sum_a \int d\Psi_a$ is
carried out over all states. According to the finite-temperature field theory, the partition function in the mean-field approximation can be expressed in the form of the path integral
\begin{equation}
	\mathcal{Z} = \int \mathcal{D} \bar{\psi} \mathcal{D}\psi \times \text{exp}\left[ \int_{0}^{\beta} d\tau \int d^3 x \mathcal{L}_{MF} \right],
\end{equation}
with the real time $t=x_0$ converted to the imaginary time $\tau = it$, and the functional integral $\int \mathcal{D} \bar{\psi} \mathcal{D}\psi $ covering all quark species. In the above equation, the condensate energy independent of $\psi$ and $\bar{\psi}$ can be factorized in $\mathcal{Z}$, and after applying the relation
\begin{eqnarray}
	\text{ln} \left(  \int \mathcal{D} i\psi^\dagger \mathcal{D}\psi e^ {i\psi^\dagger \mathcal{S}^{-1}\psi}  \right) = \text{ln det}\ \mathcal{S}^{-1}  = \text{Tr ln}\ \mathcal{S}^{-1}, \nonumber\\
\end{eqnarray}
the partition function can be simplified as
\begin{eqnarray}
	\text{ln} \mathcal{Z} &=& -i\beta V \int \frac{d^4 p}{(2\pi)^4} \text{Tr ln } \mathcal{S}(p)^{-1} -\beta V\mathcal{V} \nonumber\\
	&=& \beta V T\sum_n\int \frac{d^3 p}{(2\pi)^3} \text{Tr ln } \mathcal{S}(i\omega_n, \vec{p})^{-1} -\beta V\mathcal{V},\nonumber\\
\end{eqnarray}
where $V$ is the system volume, and the 4-momentum becomes $p = (p_0, \vec{p}) = (i\omega_n, \vec{p})$ with $\omega_n = (2n+1)\pi T$ being the Matsubara frequency for a fermi system.

The thermodynamic potential of the quark system can be obtained from the partition function through
\begin{eqnarray}\label{omega1}
	\Omega &=& -\frac{1}{\beta V} \text{ln} \mathcal{Z} \nonumber\\
	&=& -T\sum_n\int \frac{d^3 p}{(2\pi)^3} \text{Tr ln } \mathcal{S}(i\omega_n, \vec{p})^{-1} +\mathcal{V}.
\end{eqnarray}
With the form of the quark propagator as Eq.~(\ref{propagator}) and keeping in mind that $p_0=i\omega_n$, we can get the following relation after some algebras
\begin{eqnarray}
	&&\text{Trln}\mathcal{S}^{-1} (p)\nonumber\\
	&=&2N_c \,\text{ln} \Bigg\{\Big[ \left(  E_u^2-\left( p_0 + \tilde{\mu}_u \right)^2 -\Delta^2 \right) \nonumber\\
    &\times&\left( E_d^2-\left( p_0 + \tilde{\mu}_d \right)^2 - \Delta^2 \right) \nonumber\\
	&+&\left(  (M_u+M_d)^2 + 4 \vec{p}^2 - \left(2p_0 +  \tilde{\mu}_u+\tilde{\mu}_d\right)^2 \right)\Delta^2  \Big] \nonumber\\
    &\times& \left[ E_s^2-\left( p_0 + \tilde{\mu}_s \right)^2 \right]\Bigg\}, \label{3-flavor NJL determinant 1}
\end{eqnarray}
with $\Delta = \left(G_{S} + 2G_{IS} -K \sigma_s\right)\pi$ being the gap parameter, and $E_q=\sqrt{M_q^2+\vec{p}^2}$ with $q=u,d,s$ being the single-quark energy. Replacing $\tilde{\mu}_u$ and $\tilde{\mu}_d$ with the effective baryon and isospin chemical potential $\tilde{\mu}_B$ and $\tilde{\mu}_I$ according to Eq.~(\ref{eff}), Eq.~(\ref{3-flavor NJL determinant 1}) becomes
\begin{eqnarray}
	&&\text{Trln}\mathcal{S}^{-1} (p) \nonumber\\\label{3-flavor NJL determinant 2}
	&=&2N_c \,\text{ln} \Bigg\{\Big[ a\left( p_0 + \frac{\tilde{\mu}_B}{3} \right)^4 + b\left( p_0 + \frac{\tilde{\mu}_B}{3} \right)^3 \nonumber\\
    &+& c\left( p_0 + \frac{\tilde{\mu}_B}{3} \right)^2 +d \left( p_0 + \frac{\tilde{\mu}_B}{3} \right) + e\Big] \nonumber\\
    &\times& \left[ E_s^2-\left( p_0 + \tilde{\mu}_s \right)^2 \right]\Bigg\} \nonumber\\
	&=&2N_c \,\text{ln} \Bigg\{ \left[ \prod_{k=1}^4 \left(p_0 + \frac{\tilde{\mu}_B}{3} - \lambda_k\right)\right]\nonumber\\
    &\times& \left[p_0- \left( E_s- \tilde{\mu}_s \right) \right]\left[-p_0- \left( E_s + \tilde{\mu}_s \right) \right]\Bigg\},
\end{eqnarray}
where $\lambda_k$ is the solution of the quartic equation for  $p_0 + {\tilde{\mu}_B}/{3}$, with the coefficients
\begin{eqnarray}
	a &=& 1, \nonumber\\
	b &=& 0, \nonumber\\
	c &=&  -E_u^2-E_d^2 - \frac{\tilde{\mu}_I^2}{2}-2\,\Delta^2, \\
	d &=&  \tilde{\mu}_I \left(  M_u^2 - M_d^2 \right), \nonumber\\
	e &=& E_u^2E_d^2+\left(\frac{\tilde{\mu}_I}{2}\right)^4 +(E_u^2+E_d^2)\left(\Delta^2-\left(\frac{\tilde{\mu}_I}{2}\right)^2\right) \nonumber\\
      &+& \left(  \frac{\tilde{\mu}_I^2}{2} - \left( M_u-M_d\right)^2 +\Delta^2 \right)\Delta^2.\nonumber
\end{eqnarray}
In the general situation with $M_u\neq M_d$ and thus $d\neq0$, we can get the analytical expressions of the four roots $\lambda_k$
\begin{eqnarray}
	\lambda_{1} &=& +\frac{\sqrt{X}}{2} + \frac{1}{2} \sqrt{Y+\frac{Z}{4\,\sqrt{X}}}, \nonumber\\
	\lambda_{2} &=& -\frac{\sqrt{X}}{2} - \frac{1}{2} \sqrt{Y-\frac{Z}{4\,\sqrt{X}}}, \nonumber\\
	\lambda_{3} &=& -\frac{\sqrt{X}}{2} + \frac{1}{2} \sqrt{Y-\frac{Z}{4\,\sqrt{X}}}, \nonumber\\
	\lambda_{4} &=& +\frac{\sqrt{X}}{2} - \frac{1}{2} \sqrt{Y+\frac{Z}{4\,\sqrt{X}}},
\end{eqnarray}
with
\begin{equation}
X = -c+\Xi, ~~~ Y = -c-\Xi, ~~~Z = -8\,d,
\end{equation}
where $\Xi$ is expressed as
\begin{eqnarray}
	\Xi =  \frac{c}{3}+\frac{2^{\frac{1}{3}} \Xi_1}{3\,\Xi_3} + \frac{\Xi_3}{3\times2^{\frac{1}{3}}}, ~~~
\end{eqnarray}
with
\begin{eqnarray}
&&\Xi_1 = c^2+12\,e,~~~ \Xi_2 = 2\,c^3 + 27\,d^2 -72\,ce,\nonumber\\
&&\Xi_3 = \left(\Xi_2 + \sqrt{-4\Xi_1^3 + \Xi_2^2}  \right)^{\frac{1}{3}}.
\end{eqnarray}
These roots satisfy the following relations
\begin{eqnarray}
	-\sum_{k=1}^4 \lambda_k -b &=& 0, \nonumber\\
	\sum_{i<j} \prod_{k=i,j} \lambda_k -c &=& 0, \nonumber\\
	-\sum_{i=1}^4\prod_{k\neq i} \lambda_k -d &=& 0, \nonumber\\
	\prod_{k=1}^4 \lambda_k -e &=& 0.
\end{eqnarray}
In the special case of $M_u=M_d$, the four roots become
\begin{eqnarray}
	\lambda_1 &=& \sqrt{\left(E + \frac{\tilde{\mu}_I}{2}\right)^2 + \Delta^2}, \nonumber\\
	\lambda_2 &=& -\sqrt{\left(E + \frac{\tilde{\mu}_I}{2}\right)^2 + \Delta^2}, \nonumber\\
	\lambda_3 &=& \sqrt{\left(E - \frac{\tilde{\mu}_I}{2}\right)^2 + \Delta^2}, \nonumber\\
	\lambda_4 &=& -\sqrt{\left(E - \frac{\tilde{\mu}_I}{2}\right)^2 + \Delta^2},
\end{eqnarray}
with $E=\sqrt{M_u^2+\vec{p}^2}=\sqrt{M_d^2+\vec{p}^2}$. Since the sign is degenerate, there are actually only two solutions. In the baryon-free system, the general case always reduces to the special case of $M_u=M_d$.

Using the summation formula of the Matsubara frequencies
\begin{equation}
	T \sum_{n=-\infty}^{\infty}  \mathrm{ln}(i\omega_n - E)
	=\frac{E}{2}+T\mathrm{ln}(1-e^{-\beta E})\nonumber\\
\end{equation}
and combining Eqs.~(\ref{omega1}) and (\ref{3-flavor NJL determinant 2}), we can get the expression of the thermodynamic potential as Eq.~(\ref{OmegaNJL}).

\section{Quark condensate and quark density}\label{3-flavor NJL propagator pion}

In this appendix we obtain the expressions of the chiral condensates, the net-quark densities, and the pion condensate in terms of the phase-space distribution function from the quark propagator. In the absence of the pion condensate, $u$, $d$, and $s$ quarks are decoupled, and the quark propagator can be written as
\begin{eqnarray}
	\mathcal{S} =
	\begin{pmatrix}
		\mathcal{S}_{0u} &	0 		& 0
		\\
		0 	&	\mathcal{S}_{0d} 	& 0
		\\
		0 			&			 0 				& \mathcal{S}_{0s}
	\end{pmatrix}.
\end{eqnarray}
In the presence of the pion condensate, $u$ and $d$ quarks are mixed, and the off-diagonal terms appear. The quark propagator is then
\begin{eqnarray}
	\mathcal{S} =
	\begin{pmatrix}
		\mathcal{S}_{uu} &	\mathcal{S}_{ud} & 0
		\\
		\mathcal{S}_{du} &	\mathcal{S}_{dd} & 0
		\\
		0 											&			 0 			&  	\mathcal{S}_{0s}
	\end{pmatrix}\nonumber\\
\end{eqnarray}
with
\begin{eqnarray}
	\mathcal{S}_{uu}
	&=&\frac{1}{\mathcal{S}_{0u}^{-1} - \Delta\mathcal{S}_{0d}\Delta}, \nonumber\\
	\mathcal{S}_{dd}
	&=&\frac{1}{\mathcal{S}_{0d}^{-1} - \Delta\mathcal{S}_{0u}\Delta}, \nonumber\\
	\mathcal{S}_{ud}
	&=& -\mathcal{S}_{uu} \Delta \mathcal{S}_{0d} = -\mathcal{S}_{0u} \Delta \mathcal{S}_{dd}, \nonumber\\
	\mathcal{S}_{du}
	&=& -\mathcal{S}_{0d} \Delta \mathcal{S}_{uu}=-\mathcal{S}_{dd} \Delta \mathcal{S}_{0u}. \nonumber
\end{eqnarray}
In the above, the diagonal terms of the quark propagator in the absence of the pion condensate are
\begin{eqnarray}
	\mathcal{S}_{0q}({p})
	&=& \frac{ {\Lambda}_{+}^q(\vec{p})\gamma_0}{p_0-E_q^{-}} + \frac{ {\Lambda}_{-}^q(\vec{p})\gamma_0}{p_0+E_q^{+}}, \nonumber
\end{eqnarray}
with
\begin{eqnarray}
	\Lambda_{\pm}^q(\vec{p})&=&\frac{1}{2}\left[1\pm\frac{\gamma_0(\vec{\gamma} \cdot \vec{p} + M_q)}{E_q}\right], \nonumber\\
	\tilde{\Lambda}_{\pm}^q(\vec{p})&=&\frac{1}{2}\left[1\pm\frac{\gamma_0(\vec{\gamma} \cdot \vec{p} - M_q)}{E_q}\right], \nonumber
\end{eqnarray}
and $E_q^{\pm} = \sqrt{M_q^2+\vec{p}^2}\pm\tilde{\mu}_q $, for $q=u,d,s$.

The detailed expressions of each quark propagator are
\begin{eqnarray}\label{Sq}
	\mathcal{S}_{uu}({p})&=&\sum_{k=1}^4 g_{uu}\left(\lambda'_k\right)
	\frac{1}{p_0 -\lambda'_k },\nonumber\\
	\mathcal{S}_{dd} ({p})
	&=&\sum_{k=1}^4 g_{dd}\left(\lambda'_k\right)
	\frac{1}{p_0 -\lambda'_k },\nonumber\\
	\mathcal{S}_{ud}({p})
	&=&\sum_{k=1}^4 g_{ud}\left(\lambda'_k\right)
	\frac{1}{p_0 -\lambda'_k },\nonumber\\
	\mathcal{S}_{du}({p})
	&=&\sum_{k=1}^4 g_{du}\left(\lambda'_k\right)
	\frac{1}{p_0 -\lambda'_k },\nonumber\\
\end{eqnarray}
with
\begin{eqnarray}
	&&g_{uu}(\lambda'_k)\nonumber\\
	&=& \Bigg\{ \left[ \left(\lambda'_k+\tilde{\mu}_d\right)^2 - E_d^2 \right]\left(-\vec{\gamma} \cdot \vec{p} + (\lambda'_k+\tilde{\mu}_u) \gamma_0 +M_u\right) \nonumber\\
&-&\Delta^2 \left(-\vec{\gamma} \cdot \vec{p} +  (\lambda'_k+\tilde{\mu}_d)\gamma_0 +M_d\right)\Bigg\}\nonumber\\
&/&\{\prod_{j=1,j\neq k}^4 \left(\lambda'_k-\lambda'_j\right)\times \textbf{I}\},\nonumber\\
\end{eqnarray}
\begin{eqnarray}
	&&g_{dd}(\lambda'_k)\nonumber\\
	&=& \Bigg\{ \left[ \left(\lambda'_k+\tilde{\mu}_u\right)^2 - E_u^2 \right]\left(-\vec{\gamma} \cdot \vec{p} + (\lambda'_k+\tilde{\mu}_d) \gamma_0 +M_d\right) \nonumber\\
&-&\Delta^2 \left(-\vec{\gamma} \cdot \vec{p} +  (\lambda'_k+\tilde{\mu}_u)\gamma_0 +M_u\right)\Bigg\}\nonumber\\
&/&\{\prod_{j=1,j\neq k}^4 \left(\lambda'_k-\lambda'_j\right)\times \textbf{I}\},\nonumber\\
\end{eqnarray}
\begin{eqnarray}
	&&g_{ud}(\lambda'_k) \nonumber\\
	&=& \Bigg\{ \bigg[ [(\tilde{\mu}_u -\tilde{\mu}_d)\gamma_0+M_u-M_d]\vec{\gamma} \cdot \vec{p} \nonumber\\
&+& [M_d(\lambda'_k+\tilde{\mu}_u) -M_u(\lambda'_k+\tilde{\mu}_d) ]\gamma_0 \nonumber\\
&+& (\lambda'_k+\tilde{\mu}_u)(\lambda'_k+\tilde{\mu}_d)-p^2-M_uM_d\bigg] i \Delta\gamma_5  - i \Delta^3 \gamma_5\Bigg\}\nonumber\\
&/&\{\prod_{j=1,j\neq k}^4 \left(\lambda'_k-\lambda'_j\right)\times \textbf{I}\},\nonumber\\
\end{eqnarray}
\begin{eqnarray}
	&&g_{du}(\lambda'_k) \nonumber\\
	&=& \Bigg\{ \bigg[ [(\tilde{\mu}_d -\tilde{\mu}_u)\gamma_0+M_d-M_u]\vec{\gamma} \cdot \vec{p} \nonumber\\
&+& [M_u(\lambda'_k+\tilde{\mu}_d) -M_d(\lambda'_k+\tilde{\mu}_u) ]\gamma_0 \nonumber\\
&+& (\lambda'_k+\tilde{\mu}_d)(\lambda'_k+\tilde{\mu}_u)-p^2-M_dM_u\bigg] i \Delta\gamma_5  - i \Delta^3 \gamma_5\Bigg\} \nonumber\\
&/&\{\prod_{j=1,j\neq k}^4 \left(\lambda'_k-\lambda'_j\right)\times \textbf{I}\}.\nonumber
\end{eqnarray}
With the following relations \cite{HeLY-ZhuangPF2005}
\begin{eqnarray}
	\sigma_u &=& -N_c \int \frac{d^4 p}{(2\pi)^4}\mathrm{Tr}\left[i\mathcal{S}_{uu}(p) \right] \nonumber\\ &=&
	N_c\, \int \frac{d^3 p}{(2\pi)^3}T\sum_{n} \mathrm{Tr}\left[\mathcal{S}_{uu}(i\omega_n, \vec{p}) \right],  \nonumber
\end{eqnarray}
\begin{eqnarray}
	\sigma_d &=& -N_c \int \frac{d^4 p}{(2\pi)^4}\mathrm{Tr}\left[i\mathcal{S}_{dd}(p) \right] \nonumber\\ &=&
	N_c\, \int \frac{d^3 p}{(2\pi)^3}T\sum_{n} \mathrm{Tr}\left[\mathcal{S}_{dd}(i\omega_n, \vec{p}) \right],  \nonumber
\end{eqnarray}
\begin{eqnarray}
	\sigma_s &=& -N_c \int \frac{d^4 p}{(2\pi)^4}\mathrm{Tr}\left[i\mathcal{S}_{0s}(p) \right] \nonumber\\ &=&
	N_c\, \int \frac{d^3 p}{(2\pi)^3}T\sum_{n} \mathrm{Tr}\left[\mathcal{S}_{0s}(i\omega_n, \vec{p}) \right],  \nonumber
\end{eqnarray}
\begin{eqnarray}
	\rho_u &=& -N_c \int \frac{d^4 p}{(2\pi)^4}\mathrm{Tr}\left[i\mathcal{S}_{uu}(p)\gamma_0 \right] \nonumber\\ &=&
	N_c\, \int \frac{d^3 p}{(2\pi)^3}T\sum_{n} \mathrm{Tr}\left[\mathcal{S}_{uu}(i\omega_n, \vec{p})\gamma_0 \right],  \nonumber
\end{eqnarray}
\begin{eqnarray}
	\rho_d &=& -N_c \int \frac{d^4 p}{(2\pi)^4}\mathrm{Tr}\left[i\mathcal{S}_{dd}(p)\gamma_0 \right] \nonumber\\ &=&
	N_c\, \int \frac{d^3 p}{(2\pi)^3}T\sum_{n} \mathrm{Tr}\left[\mathcal{S}_{dd}(i\omega_n, \vec{p})\gamma_0 \right],  \nonumber
\end{eqnarray}
\begin{eqnarray}
	\rho_s &=& -N_c \int \frac{d^4 p}{(2\pi)^4}\mathrm{Tr}\left[i\mathcal{S}_{0s}(p)\gamma_0 \right] \nonumber\\ &=&
	N_c\, \int \frac{d^3 p}{(2\pi)^3}T\sum_{n} \mathrm{Tr}\left[\mathcal{S}_{0s}(i\omega_n, \vec{p})\gamma_0 \right],  \nonumber
\end{eqnarray}
\begin{eqnarray}
	\pi    &=& N_c \int \frac{d^4 p}{(2\pi)^4}\mathrm{Tr}\left[\mathcal{S}_{ud}(p)\gamma_5 + \mathcal{S}_{du}(p)\gamma_5\right] \nonumber\\ &=&
	N_c\, \int \frac{d^3 p}{(2\pi)^3}T\sum_{n} \mathrm{Tr}\left[i\mathcal{S}_{ud}(i\omega_n, \vec{p})\gamma_5 + i\mathcal{S}_{du}(i\omega_n, \vec{p})\gamma_5 \right],  \nonumber
\end{eqnarray}
and the expressions of the quark propagators [Eq.~(\ref{Sq})], we can obtain the analytical expressions of the chiral condensates, the net-quark densities, and the pion condensate as Eqs.~(\ref{gap_equation1})-(\ref{gap_equation6}) and Eq.~(\ref{pi}).

\begin{acknowledgments}

JX was supported by the National Natural Science Foundation of China under Grant No. 11922514. GXP was supported by the National Natural Science Foundation of China under Grant Nos. 11875052, 11575190, and 11135011.

\end{acknowledgments}


\begin{thebibliography}{99}

%LQCD
\bibitem{Kar02} F. Karsch, Lect. Notes Phys. \textbf{583}, 209 (2002).

\bibitem{Mur03} S. Muroya, A. Nakamura, C. Nonaka, and T. Takaishi, Prog. Theor. Phys. \textbf{110}, 615 (2003).

\bibitem{Ber05} C. Bernard {\it et al.}, Phys. Rev. D \textbf{71}, 034504 (2005). %sign problem

\bibitem{Aok06} Y. Aoki {\it et al.}, Nature \textbf{443}, 675 (2006). %sign problem

\bibitem{Baz12a} A. Bazavov {\it  et al.}, Phys. Rev. D \textbf{85}, 054503 (2012). %sign problem

\bibitem{Bed18} P. F. Bedaque, EPJ Web Conf. \textbf{175}, 01020 (2018).

%NJL
\bibitem{Bra13} N. M. Bratovic, T. Hatsuda, and W. Weise, Phys. Lett. B \textbf{719}, 131 (2013).

\bibitem{Asa89} M. Asakawa and K. Yazaki, Nucl. Phys. A \textbf{504}, 668 (1989).

\bibitem{Fuk08} K. Fukushima, Phys. Rev. D \textbf{77}, 114028 (2008); {\it ibid.} \textbf{78}, 039902 (E) (2008).

\bibitem{Car10} S. Carignano, D. Nickel, and M. Buballa, Phys. Rev. C \textbf{82}, 054009 (2010).

%DS approach
\bibitem{Xin14} X. Y. Xin, S. X. Qin, and Y. X. Liu, Phys. Rev. D \textbf{90}, 076006 (2014).

\bibitem{Fis14} C. S. Fischer, J. Luecker, and C. A. Welzbacher, Phys. Rev. D \textbf{90}, 034022 (2014).

%FRQ approach
\bibitem{Fu19} W. J. Fu, J. M. Pawlowski, and F. Rennecke, Phys. Rev. D \textbf{101}, 054032 (2020).

\bibitem{Gao20} F. Gao and J. M. Pawlowski, Phys. Rev. D \textbf{102}, 034027 (2020).

%pion condensate
\bibitem{Son01} D. T. Son and M. A. Stephanov, Phys. Rev. Lett. \textbf{86}, 592 (2001).

\bibitem{Kle03} B. Klein, D. Toublan, and J. J. M. Verbaarschot, Phys. Rev. D \textbf{68}, 014009 (2003).

\bibitem{Bar04} A. Barducci, R. Casalbuoni, G. Pettini, and L. Ravagli, Phys. Rev. D \textbf{69}, 096004 (2004).

\bibitem{HeLY-ZhuangPF2005} L. He, M. Jin, and P. Zhuang, Phys. Rev. D \textbf{71}, 116001 (2005).

\bibitem{Bar05} A. Barducci, R. Casalbuoni, G. Pettini, and L. Ravaglix, Phys. Rev. D \textbf{72}, 056002 (2005).

\bibitem{Ebe06a} D.~Ebert and K.~G.~Klimenko,
%``Gapless pion condensation in quark matter with finite baryon density,''
J. Phys. G \textbf{32}, 599 (2006).

\bibitem{Zha07} Z. Zhang and Y. X. Liu, Phys. Rev. C \textbf{75}, 035201 (2007).

\bibitem{Zha07b} Z. Zhang and Y. X. Liu, Phys. Rev. C \textbf{75}, 064901 (2007).

\bibitem{Sas10} T. Sasaki, Y. Sakai, H. Kouno, and M. Yahiro, Phys. Rev. D \textbf{82}, 116004 (2010).

\bibitem{Fu10} M. C. Fu, L. Y. He, and Y. X. Liu, Phys. Rev. D \textbf{82}, 056006 (2010).

\bibitem{Xia13} T. Xia, L. Y. He, and P. F. Zhuang, Phys. Rev. D \textbf{88}, 056013 (2013).

\bibitem{Adh18} P. Adhikari, J. O. Andersen, and P. Kneschke, Phys. Rev. D \textbf{98}, 074016 (2018).

\bibitem{Bra18a} B. B. Brandt, G. Endr\"odi, and S. Schmalzbauer, Phys. Rev. D \textbf{97}, 054514 (2018).

\bibitem{Khu19} T. G. Khunjua, K. G. Klimenko, and R. N. Zhokhov, Phys. Rev. D \textbf{100}, 034009 (2019).

\bibitem{Wu21} Z. Q. Wu, J. L. Ping, and H. S. Zong, Chin. Phys. C \textbf{45}, 064102 (2021).

%isovector interaction in NJL
\bibitem{Liu19} H. Liu, F. T. Wang, K. J. Sun, J. Xu, and C. M. Ko, Phys. Lett. B \textbf{798}, 135002 (2019).

\bibitem{Liu20a} H. Liu and J. Xu, Universe \textbf{7}, 6 (2021).

\bibitem{Chu15} P. C. Chu, X. Wang, L. W. Chen, and M. Huang, Phys. Rev. D \textbf{91}, 023003 (2015).

\bibitem{Liu20} H. Liu, J. Xu, and C. M. Ko, Phys. Lett. B \textbf{803}, 135343 (2020).

%fit LQCD
\bibitem{Ava19} S. S. Avancini, A. Bandyopadhyay, D. C. Duarte, and R. L. S. Farias, Phys. Rev. D \textbf{100}, 116002 (2019). %NJL fit cold LQCD results

\bibitem{Lop21} B. S. Lopes, S. S. Avancini, A. Bandyopadhyay, D. C. Duarte, and R. L. S. Farias, Phys. Rev. D \textbf{103}, 076023 (2021). %NJL fit hot LQCD results

\bibitem{Cao21} G. Q. Cao, P. F. Zhuang, and L. Y. He, arXiv: 2105.08932 [hep-ph].

%NJL
\bibitem{Liu16} H. Liu, J. Xu, L. W. Chen, and K. J. Sun, Phys. Rev. D \textbf{94}, 065032 (2016).

\bibitem{KMT} G. 't Hooft, Phys. Rev. D \textbf{14}, 3432 (1976); Phys. Rev. D \textbf{18}, 2199 (1978).

\bibitem{Lut92} M. Lutz, S. Klimt, and W. Weise, Nucl. Phys. A \textbf{542}, 521 (1992).

\bibitem{He05} L. Y. He and P. F. Zhuang, Phys. Lett. B \textbf{615}, 93 (2005).

%LQCD
\bibitem{Bra18} B. B. Brandt, G. Endr\"odi, E. S. Fraga, M. Hippert, J. Schaffner-Bielich, and S. Schmalzbauer, Phys. Rev. D \textbf{98}, 094510 (2018).

\bibitem{Det12} W. Detmold, K. Orginos, and Z. Shi, Phys. Rev. D \textbf{86}, 054507 (2012).

\bibitem{Ebe06} D. Ebert and K. G. Klimenko, Eur. Phys. J. C \textbf{46}, 771 (2006).

\bibitem{Ade10} J.~O.~Andersen and L.~Kyllingstad,
%``Pion Condensation in a two-flavor NJL model: the role of charge neutrality,''
J. Phys. G \textbf{37}, 015003 (2010).

%sarma phase
\bibitem{He06} L. Y. He, M. Jin, and P. F. Zhuang, Phys. Rev. D \textbf{74}, 036005 (2006).

\bibitem{For05} M. M. Forbes, E. Gubankova, W. V. Liu, and F. Wilczek, Phys. Rev. Lett. \textbf{94}, 017001 (2005).

\bibitem{sar63} G. Sarma, J. Phys. Chem. Solids \textbf{24}, 1029 (1963).

\bibitem{Mu09} C. F. Mu and P. F. Zhuang, Phys. Rev. D \textbf{79}, 094006 (2009).

\bibitem{Boe15} I. Boettcher, T. K. Herbst, J. M. Pawlowski, N. Strodthoff, L. von Smekal, and C. Wetterich, Phys. Lett. B \textbf{742}, 86 (2015).

%chiral imbalance
\bibitem{Khu20} T. G. Khunjua, K. G. Klimenko, and R. N. Zhokhov, Eur. Phys. J. C \textbf{80}, 995 (2020).

%color superconductivity
\bibitem{Sun07} G. F. Sun, L. Y. He, and P. F. Zhuang, Phys. Rev. D \textbf{75}, 096004 (2007).

\bibitem{Mu10} C. F. Mu, L. Y. He, and Y. X. Liu, Phys. Rev. D \textbf{82}, 056006 (2010).

\end{thebibliography}
\end{document}